%% file: 2dnc_04.tex
\documentclass[12pt]{article}
\usepackage{epsfig,multicol,multirow,cite}
\usepackage{amsmath,subfigure,latexsym,amssymb}
\usepackage{graphicx}
\usepackage{collref}
\usepackage{bbm}
\usepackage{color}
\usepackage{hyperref}
\def\lsi{\raise0.3ex\hbox{$<$\kern-0.75em\raise-1.1ex\hbox{$\sim$}}}
\def\gsi{\raise0.3ex\hbox{$>$\kern-0.75em\raise-1.1ex\hbox{$\sim$}}}
\def\backder{\raise1.4ex\hbox{$\leftarrow$\kern-0.75em\raise-1.4ex\hbox{$\partial$}}}
\def\forder{\raise1.4ex\hbox{$\rightarrow$\kern-0.75em\raise-1.4ex\hbox{$\partial$}}}
\newcommand{\backderi}{\mathop{\backder}}
\newcommand{\forderi}{\mathop{\forder}}

\newcommand{\be}{\begin{equation}}
\newcommand{\ee}{\end{equation}}
\newcommand{\nn}{\nonumber}
\newcommand{\bea}{\begin{eqnarray}}
\newcommand{\eea}{\end{eqnarray}}

\newcommand{\la}{\langle}
\newcommand{\ra}{\rangle}

\newcommand{\R}{{\kern+.25em\sf{R}\kern-.78em\sf{I} \kern+.78em\kern-.25em}}
\newcommand{\N}{{\kern+.25em\sf{N}\kern-.78em\sf{I} \kern+.78em\kern-.25em}}
\newcommand{\C}{{\kern+.25em\sf{C}\kern-.50em\sf{I} \kern+.50em\kern-.25em}}

\newcommand{\ri}{{\rm i}}

\makeatletter
\@addtoreset{equation}{section}
\makeatother

\begin{document}

\begin{flushright}
IFT-UAM/CSIC-14-014
\end{flushright}
\vspace*{1mm}

\begin{center}

{\Large{\bf The Continuum Phase Diagram of the}} \\

\vspace*{4mm}

{\Large\bf 2d Non-Commutative $\lambda \phi^{4}$ Model} 

\vspace*{1cm}
 
H\'{e}ctor Mej\'{\i}a-D\'{\i}az$^{\rm \, a}$, Wolfgang Bietenholz$^{\rm \, a}$
and Marco Panero$^{\rm \, b}$  \\

\vspace*{8mm}

{\small
$^{\rm a}$ Instituto de Ciencias Nucleares \\
Universidad Nacional Aut\'{o}noma de M\'{e}xico \\
A.\ P.\ 70-543, C.\ P.\ 04510 Distrito Federal, Mexico

\vspace*{4mm}

$^{\rm b}$ Instituto de F\'{\i}sica Te\'{o}rica UAM/CSIC \\ 
Universidad Aut\'{o}noma de Madrid \\
Ciudad Universitaria de Cantoblanco \\
28049 Madrid, Spain
}
\end{center}

\vspace*{8mm}

\input{abs}

\newpage

\section{Introduction}

\input{intro}

\section{Formulation on the lattice and as a matrix model}
\label{sec:matmod}
\input{matmod}

\section{The phase diagram on the lattice}
\label{sec:phasedia}
\input{phasedia2}

\section{The Double Scaling Limit}
\label{sec:DSL}
\input{DSL2}

\section{Conclusions}
\label{sec:conclu}
\input{conclu2}

\input{ackno}

\appendix

\section{Numerical techniques}
\label{app:numtech}
\input{numtech}

\end{document}

%% file: abs.tex
We present a non-perturbative study of the $\lambda \phi^{4}$ 
model on a non-commu\-ta\-tive plane. The lattice regularised form
can be mapped onto a Hermitian matrix model, which enables 
Monte Carlo simulations. Numerical data reveal the phase 
diagram; at large $\lambda$ it contains a ``striped phase'',
which is absent in the commutative case.
We explore the question whether or not this phenomenon 
persists in a Double Scaling Limit (DSL), which extra\-polates 
simultaneously to the continuum and to infinite volume, 
at a fixed non-commutativity parameter. To this end, we introduce 
a dimensional lattice spacing based on the decay of the correlation 
function. Our results provide evidence for the existence of a 
striped phase even in the DSL, which implies the spontaneous 
breaking of translation symmetry. Due to the non-locality of 
this model, this does not contradict the Mermin-Wagner theorem.

%% file: intro.tex
Since the dawn of the new millennium, field theory on non-commutative 
spaces (``NC field theory'') has attracted a lot of interest, see Refs.\
\cite{DougNek,Szabo} for reviews. The idea as such is historic \cite{Sny}, 
but the observation that it can be related to low energy string 
theory \cite{SeiWit} triggered an avalanche of thousands of
papers on this subject. 

The point of departure are space-time coordinates, which
are given by Hermitian operators that do not commute.
In the functional integral approach, however,
NC field theory can be formulated with ordinary
space-time coordinates $x$, if the field
multiplications are carried out using the Moyal
star-product,
\begin{equation}
\phi (x) \star \psi (x) : = \phi (x) \exp \Big( \frac{\ri}{2}
\backderi \,\! _{\mu} \,
\Theta_{\mu \nu} \forderi \,\! _{\nu} \, \Big) \psi (x) \ .
\end{equation}
In particular the star-commutator of the coordinates,
\begin{equation}  \label{NCrelat}
x_{\mu} \star x_{\nu} - x_{\nu} \star x_{\mu} = \ri \Theta_{\mu \nu} \ ,
\end{equation}
yields the (real, anti-symmetric) non-commutativity ``tensor'' $\Theta$. 
Here we consider the simplest case, where $\Theta$ is constant.

The perturbative treatment of NC field theory is obstructed by the 
notorious problem of mixing between ultraviolet and infrared (UV/IR)
degrees of freedom, {\it i.e.}\ the appearance of 
singularities at both ends of the energy scale in non-planar 
diagrams \cite{UVIR}. Therefore, renormalisation beyond one loop 
is mysterious. Effects due to this mixing also show up at the 
non-perturbative level, as this study is going to confirm.

In quadratic, integrated terms (with vanishing boundary contributions)
the star-product is equivalent to the ordinary product. Hence the 
Euclidean action of the NC $\lambda \phi^{4}$ model takes the 
form\footnote{As usual, the energy dimension of $\phi$ is $(d-2)/2$ 
and the dimension of the quartic coupling $\lambda$ is $(4-d)$.}
\begin{equation}
S [ \phi ] = \int d^{d}x \, \Big[ \frac{1}{2}
\partial_{\mu} \phi \, \partial_{\mu} \phi + \frac{m^{2}}{2} \phi^{2}
+ \frac{\lambda}{4} \phi \star \phi \star \phi \star \phi \Big] \ ,
\end{equation}
so $\lambda$ does not only control the strength of the coupling, 
but also the extent of the non-commutativity effects. 
 
A 1-loop calculation in the framework of a Hartree-Fock
approach led to the following conjecture about the phase diagram
of this model in $d = 3$ and $4$ \cite{Gubs}:
there is a disordered phase, and --- at strongly negative $m^{2}$ --- 
an order regime. The latter splits into a phase of uniform magnetisation
(at small $\lambda$, with an Ising-type transition to disorder, as in 
the commutative case), and a ``striped phase'' (at large $\lambda$),
where periodic non-uniform magnetisation patterns dominate. 
That phase, which does not exist in the commutative 
$\lambda \phi^{4}$ model, was also discussed based on renormalisation 
group methods \cite{Wu} and on the Cornwall-Jackiw-Tomboulis 
effective action \cite{Zap}. In $d=3$, more precisely in the 
case of a NC plane and a commutative Euclidean time direction, 
the existence of such a striped phase was observed 
explicitly in a non-perturbative, numerical study on the lattice 
\cite{BHN}, in agreement with the qualitative prediction of 
Ref.\ \cite{Gubs}.

The numerical computation of the dispersion 
relation $E = E ( \vec p )$ (with $\vec p = (p_{1},p_{2})$) 
revealed that for $\lambda$ values above some critical $\lambda_{\rm c}$,
the energy $E$ grows not only at large momenta, but also
at very small $| \vec p |$, 
hence $E$ takes its minimum at some finite momentum. 
A strongly negative parameter $m^{2}$ enforces
ordering with the pattern corresponding to the modes of minimal energy:
this results in a (Poincar\'e symmetry breaking) type of order
which can be defined as a ``stripe pattern''.
The momentum value corresponding to the minimum $E$ stabilises as 
one approaches a Double Scaling Limit (DSL), {\it i.e.}\ 
by taking the continuum (lattice spacing $a \to 0$) and 
thermodynamic (physical volume and number of degrees of freedom
tending to infinity) limits,
at fixed non-commutativity tensor \cite{BHN}. This shows that the 
striped phase does indeed exist in $d=3$, which is a manifestation of 
the coexistence of UV and IR divergences. 
In addition, it also suggests non-perturbative renormalisability.

Such a striped phase was also observed numerically in the 2d lattice model 
\cite{BHN,AmbCatt}, but the survival of that 
phase in the DSL has never been clarified.
One might suspect that it does not survive this limit, 
{\it i.e.}\ that the DSL removes the phase boundary inside
to ordered regime, $\lambda_{\rm c} \to \infty$, because in such a 
phase translation and rotation symmetry are spontaneously broken. 
Indeed, in lattice field theory there exist phases, which are
regularisation artifacts, such as the confinement
phase in lattice Quantum Electrodynamics 
at strong coupling. Na\"{\i}vely, one could think that the persistence 
of the striped phase in the continuum limit would violate the
Mermin-Wagner theorem \cite{MW}, which rules out the spontaneous breaking of 
continuous, global symmetries in $d \leq 2$. However,
this theorem is based on assumptions like locality and a regular IR 
behaviour, which do not hold here, and hence it is not
automatically applicable in NC field theory.

Therefore Ref.\ \cite{Gubs} presented a refined consideration, 
and conjectured the absence of a striped phase in $d=2$.
The authors argued based on an effective action of the Brazovskii 
form: in this formulation, the kinetic term is of 
fourth order in the momentum, which renders the statement
of the Mermin-Wagner theorem more powerful. This suggests that 
it might even capture the NC case. On the other hand, another 
effective action approach supported
the existence of a striped phase in $d=2$ \cite{Zap08}.

Here we investigate this controversial question numerically;
this has been reported before in a thesis and in a
proceeding contribution \cite{Hector}.
In Section \ref{sec:matmod} we review the matrix model formulation; in this 
form, the model is tractable by Monte Carlo simulations. Section 
\ref{sec:phasedia} describes the phase diagram on the 
lattice. Section \ref{sec:DSL} introduces
a physical scale in order to address the question if there exists
a stable DSL in the vicinity of the striped phase. We summarise
our conclusions in Section \ref{sec:conclu}. Some numerical tricks, 
which are useful in the simulation of Hermitian matrix models,
are sketched in Appendix \ref{app:numtech}.

%% file: matmod.tex
As relation (\ref{NCrelat}) shows, the points on a non-commutative plane are
somewhat washed out, hence it is not possible to discretise a 
non-commutative plane by means of a lattice with sharp sites.
On the other hand, the momentum components are assumed to
commute with each other. As suggested in Refs.~\cite{Szabo,AMNS}, a 
(fuzzy) lattice structure can be introduced, by restricting the momenta 
of the NC theory to the first Brillouin zone. This is only consistent 
for discrete momenta, hence the NC lattice is automatically periodic.
In particular, on a periodic $N\times N$ lattice of spacing $a$, 
the non-commutativity parameter $\theta$ is identified as
\begin{equation}  \label{thetaNa2}
\theta = \frac{1}{\pi} N a^{2} \ , \qquad {\rm with} \quad
\Theta_{\mu \nu} = \theta \epsilon_{\mu \nu} \ .
\end{equation}
The DSL consists of the simultaneous limits 
$N \to \infty$ and $a \to 0$ at $N a^{2} = {\rm const.}$, 
{\it i.e.}\ one extrapolates to the continuum and to infinite 
volume (since $Na$ also diverges), while keeping the
non-commutativity parameter $\theta$ constant.

However, such a formulation is still unpractical for numerical
simulations, because the star-product couples field variables
defined on {\em any} pair of lattice sites. 
Fortunately, a computer-friendly formulation 
can be obtained by mapping the system to a twisted matrix model. 
Such a mapping was first suggested in the context of NC $U(N)$ 
gauge theory \cite{AIIKKT} (in which twisted boundary conditions are 
required for the algebra to close).\footnote{Large-$N$ Yang-Mills 
theories defined on a space with twisted boundary conditions were 
first introduced in Ref.~\cite{TEK} and used in simulations of
NC gauge theories \cite{NCgauge}; see also Subsection 4.7 of 
Ref.~\cite{Lucini:2012gg} for an introduction to this subject, 
and Ref.~\cite{Marga} for recent applications.}
Similarly, the matrix formulation of the NC $\lambda \phi^{4}$ model 
takes the form \cite{AMNS}
\begin{equation}  \label{matmod}
S [ \Phi ] = N {\rm Tr} \left[ \frac{1}{2}
\sum_{\mu = 1}^{2} \Big( \Gamma_{\mu} \Phi \Gamma_{\mu}^{\dagger}
- \Phi \Big)^{2} + \frac{\bar m^{2}}{2} \Phi^{2}
+ \frac{\bar \lambda}{4} \Phi^{4} \right] \ ,
\end{equation}
where $\Phi$ is a Hermitian $N \times N$ matrix, which captures
in one point all the degrees of freedom of the lattice field, 
while ${\bar m^{2}}$ and ${\bar \lambda}$ are the (dimensionless)
parameters of the matrix model, expressed in
units of the lattice spacing $a$.
The $\Gamma_{\mu}$ are unitary matrices called
{\it twist eaters}: they satisfy the 't~Hooft-Weyl algebra
\begin{equation}
\Gamma_{\mu} \Gamma_{\nu} = {\cal Z}_{\nu \mu} \Gamma_{\nu} \Gamma_{\mu} \ ,
\end{equation}
where the ${\cal Z}$ tensor encodes the twist factor in the boundary 
conditions. We choose ${\cal Z}_{12} = {\cal Z}_{21}^{*} = \exp 
\left[ \ri \pi (N + 1) / N \right]$, where the matrix (or lattice) size $N$ has 
to be odd. This choice, and the form of the twist eaters corresponding to ``shift'' and ``clock'' matrices,
\be
\vspace*{2mm}
\Gamma_{1} = \left( \begin{array}{cccccc}
0 & 1 & 0 & . & . & 0 \\
0 & 0 & 1 & . & . & . \\
0 & 0 & 0 & . & . & . \\
. & . & . & . & . & . \\
. & . & . & . & 0 & 1 \\
1 & . & . & . & 0 & 0 \end{array} \right) \quad , \quad
\Gamma_{2} = \left( \begin{array}{cccccc}
1 & 0 & 0 & . & . & ~0 \\
0 & {\cal Z}_{21} & 0 & . & . & ~. \\
0 & 0 & {\cal Z}_{21}^{2} & . & . & ~. \\
. & . & . & . & . & ~. \\
. & . & . & . & {\cal Z}_{21}^{N-2} & 0 \\
0 & . & . & . & 0 & {\cal Z}_{21}^{N-1}  \end{array} \right) \nn
\vspace*{2mm}
\ee
follows Refs.\ \cite{AMNS,BHN}, and is similar to the 
formulation in Ref.\ \cite{AmbCatt}.

This type of matrix model has also been studied in Ref.\ \cite{Stein},
which further supported the scenario of a striped phase phase 
in $d=2$ and $4$. Moreover, the formulation of
the $\lambda \phi^{4}$ model on a ``fuzzy sphere'' also leads to
NC coordinates on the regularised level, though they obey a different
non-commutativity relation, where $\Theta (x)$ is not constant.
Also that model can be translated into a Hermitian matrix 
formulation similar to eq.\ (\ref{matmod}),
which has been studied numerically in $d=2$ \cite{Xav} and in $d=3$ 
\cite{Jul}. As a further variant, Ref.\ \cite{DGGM} studied 
the $\lambda \phi^{4}$ model on a fuzzy cylinder.
In all these cases, the phase diagram at finite $N$ 
involves a striped phase as well.

%% file: phasedia2.tex
\begin{figure}[htbp]
\vspace*{-4mm}
  \centering
  \includegraphics[width=1.\linewidth]{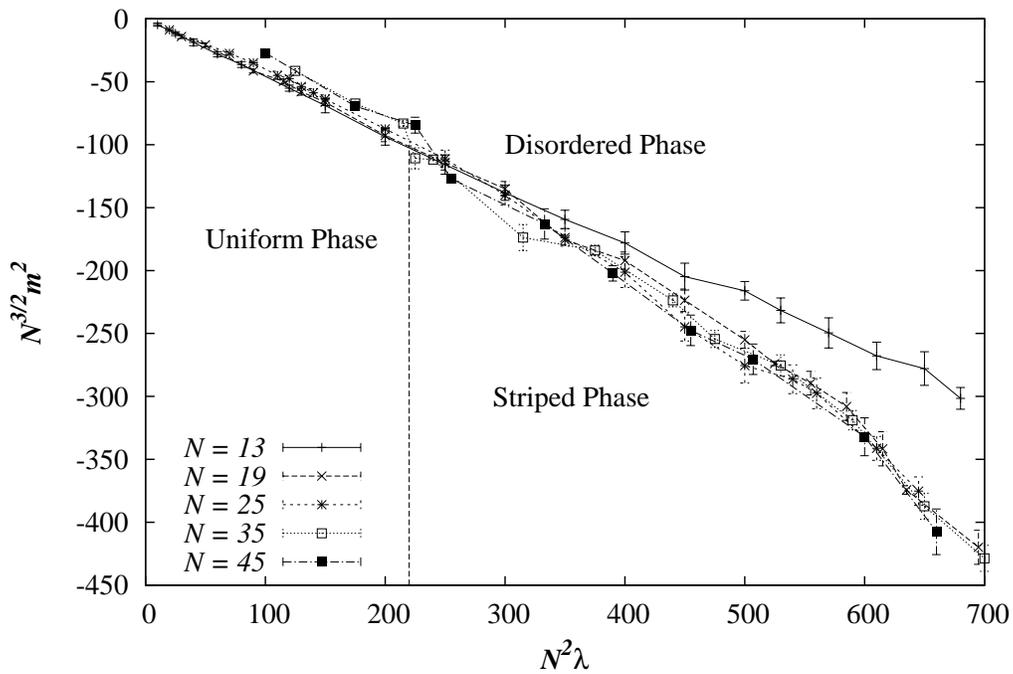}
\vspace*{-4mm}
\caption{The phase diagram of the 2d NC $\lambda \phi^{4}$ model.
The lattice size is $N \times N$, and a stable
phase transition line between order and disorder is observed for $N \geq 19$.
At small coupling $\lambda$ this order is uniform, but at
large $\lambda$, where NC effects are no longer negligible, it becomes 
striped. (The quantities on both axes are expressed in units of $a^{-2}$.)}
\label{phasedia}
\end{figure}

Figure \ref{phasedia} shows the phase diagram obtained
from our Monte Carlo simulations of this model, in its matrix
formulation (\ref{matmod}).\footnote{For simplicity, throughout this 
section all dimensional quantities are expressed in lattice units.}
The qualitative features of the phase diagram agree 
with those in $d=3$ \cite{Gubs}. There is a 
disordered phase at positive or weakly negative $m^{2}$; when 
this parameter decreases below some critical value 
$m_{\rm c}^{2} (\lambda) < 0$, the disordered phase gives
way to an order regime. At small $\lambda$ this order is uniform, 
but at larger $\lambda$ it is ``striped'', {\it i.e.}\ the ground 
state breaks the rotational and translational symmetries 
of the system. The precise location of the
boundary between the uniform- and non-uniform-order phases 
is particularly challenging to identify, hence the dotted vertical
line in Figure \ref{phasedia} is only suggestive.

On the other hand, the boundary between the disordered and the ordered
phases can be identified well, as described below. To a very good 
accuracy, its location stabilises
for $N \gtrsim 19$, if the phase diagram is plotted as a function of
$N^{3/2}m^{2}$ and $N^{2} \lambda$. Note that this differs
from the 3d case, where the choice of axes making the scaling 
properties manifest is $N^{2}m^{2}$ and $N^{2} \lambda$ \cite{BHN}.
In $d=2$, in the limit of very large quartic couplings --- in which 
the kinetic term (in the matrix formulation) becomes negligible --- 
the theory reduces to a one-matrix model, for which the boundary of 
the disordered phase is given by $\bar m_{\rm c}^{2} 
= -2 \sqrt{\bar \lambda}$ \cite{1matrix}. This behaviour 
was confirmed numerically in the related fuzzy sphere model 
\cite{Xav}, but it is not visible in Figure \ref{phasedia}, which
does not extend to very large values of $\lambda$.

\begin{figure}[hbt!]
\center
\includegraphics[angle=0,width=0.5\linewidth]{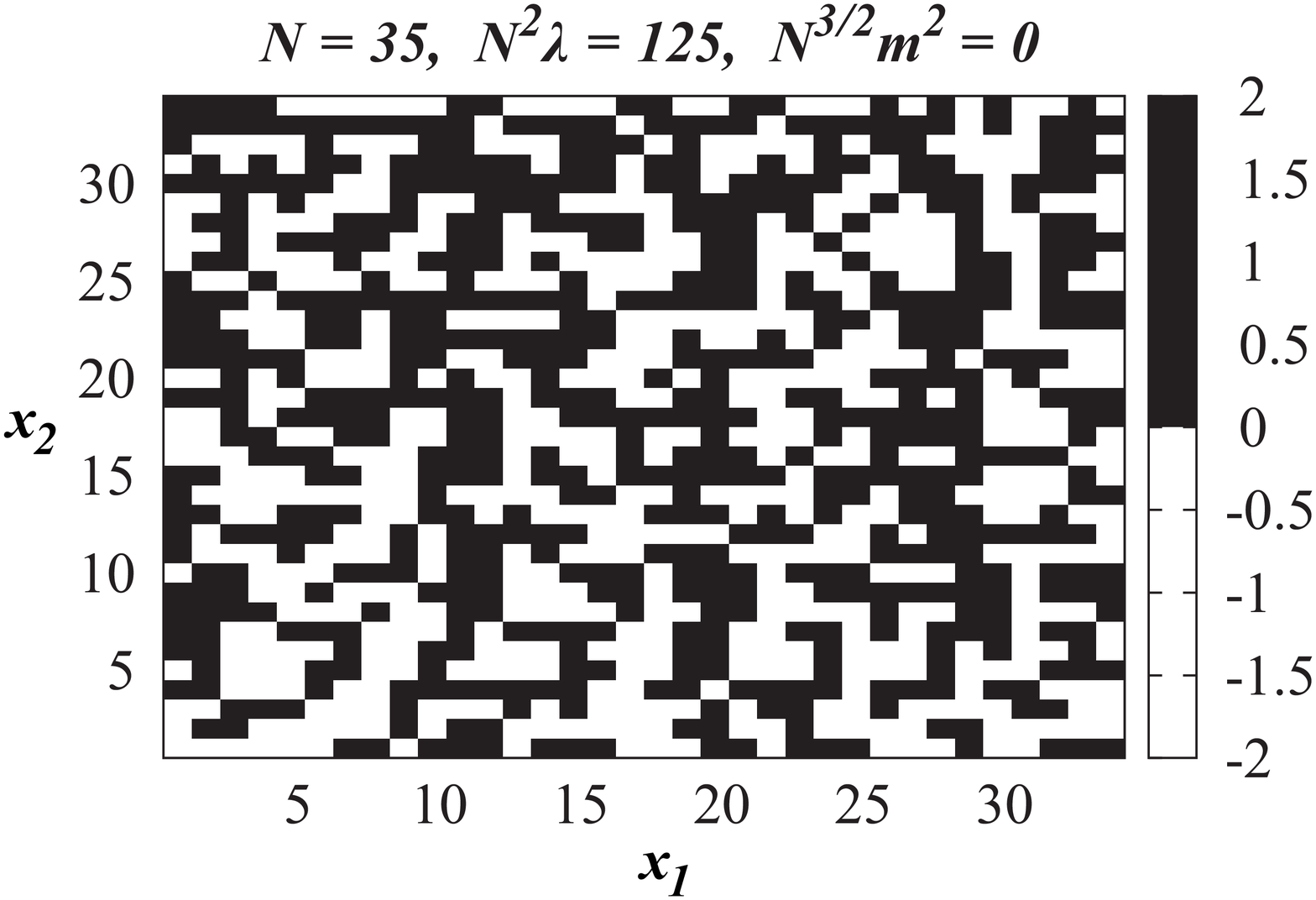}
\hspace{-3mm}
\includegraphics[angle=0,width=0.5\linewidth]{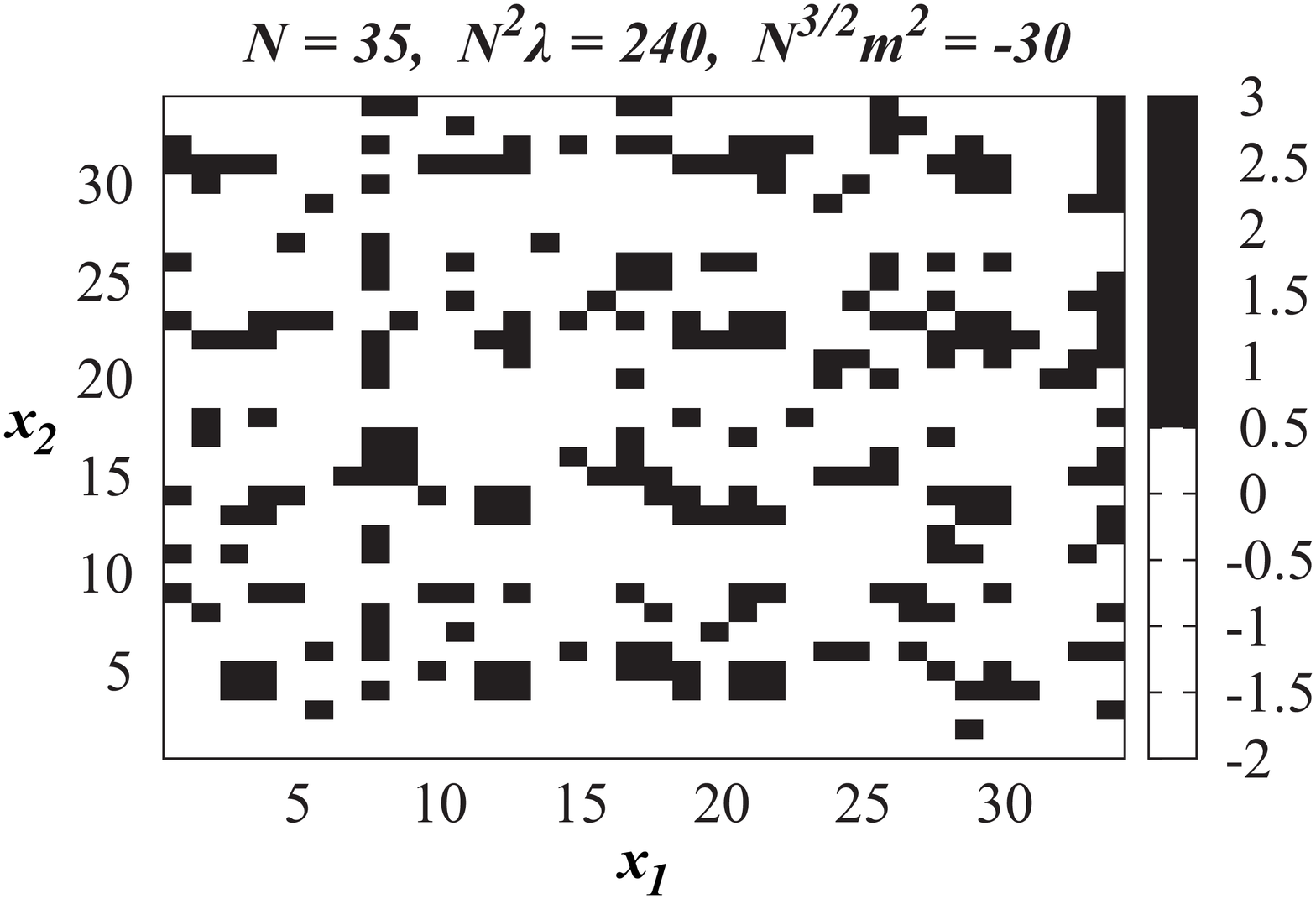} \\
\includegraphics[angle=0,width=0.5\linewidth]{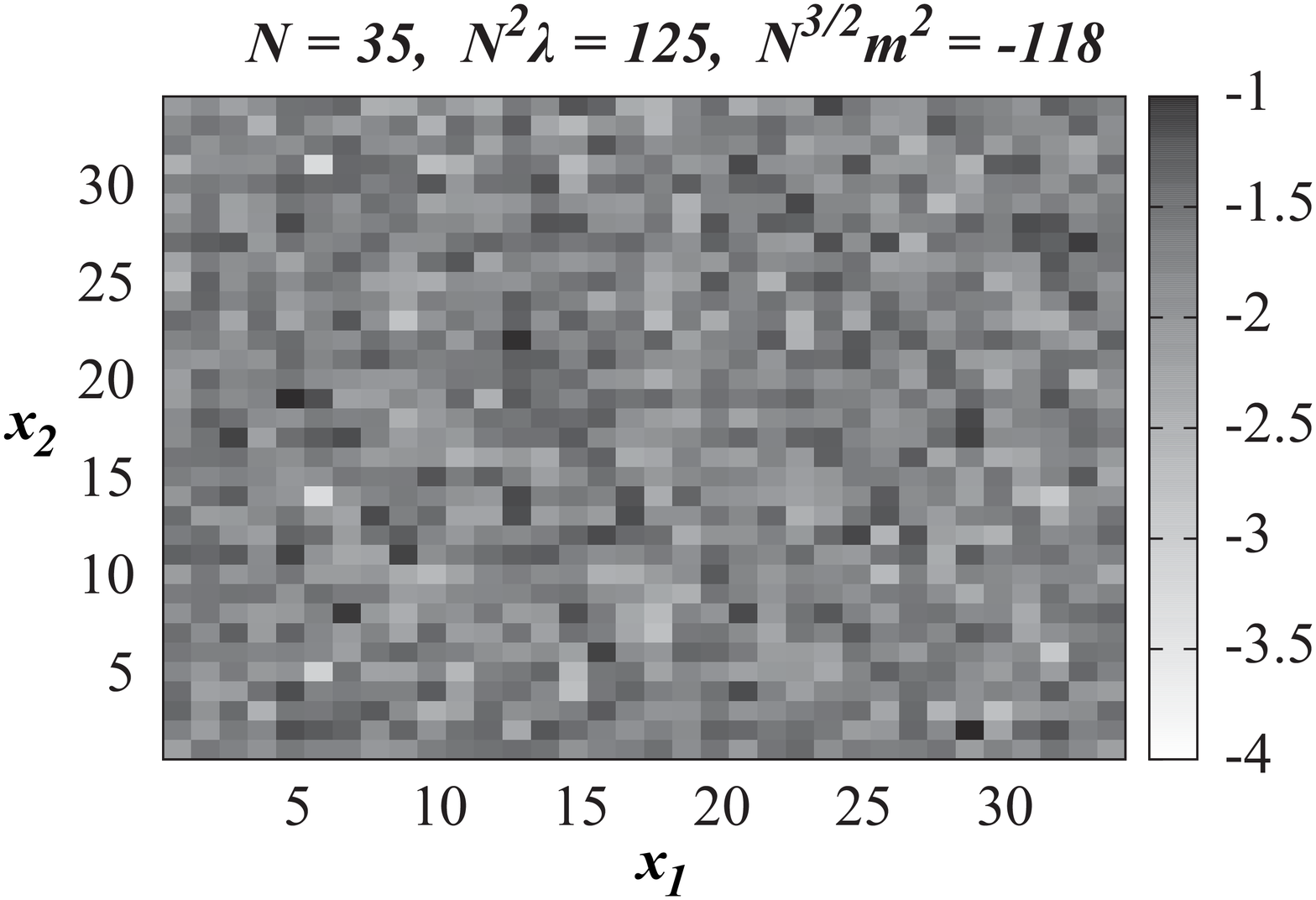}
\hspace{-3mm}
\includegraphics[angle=0,width=0.5\linewidth]{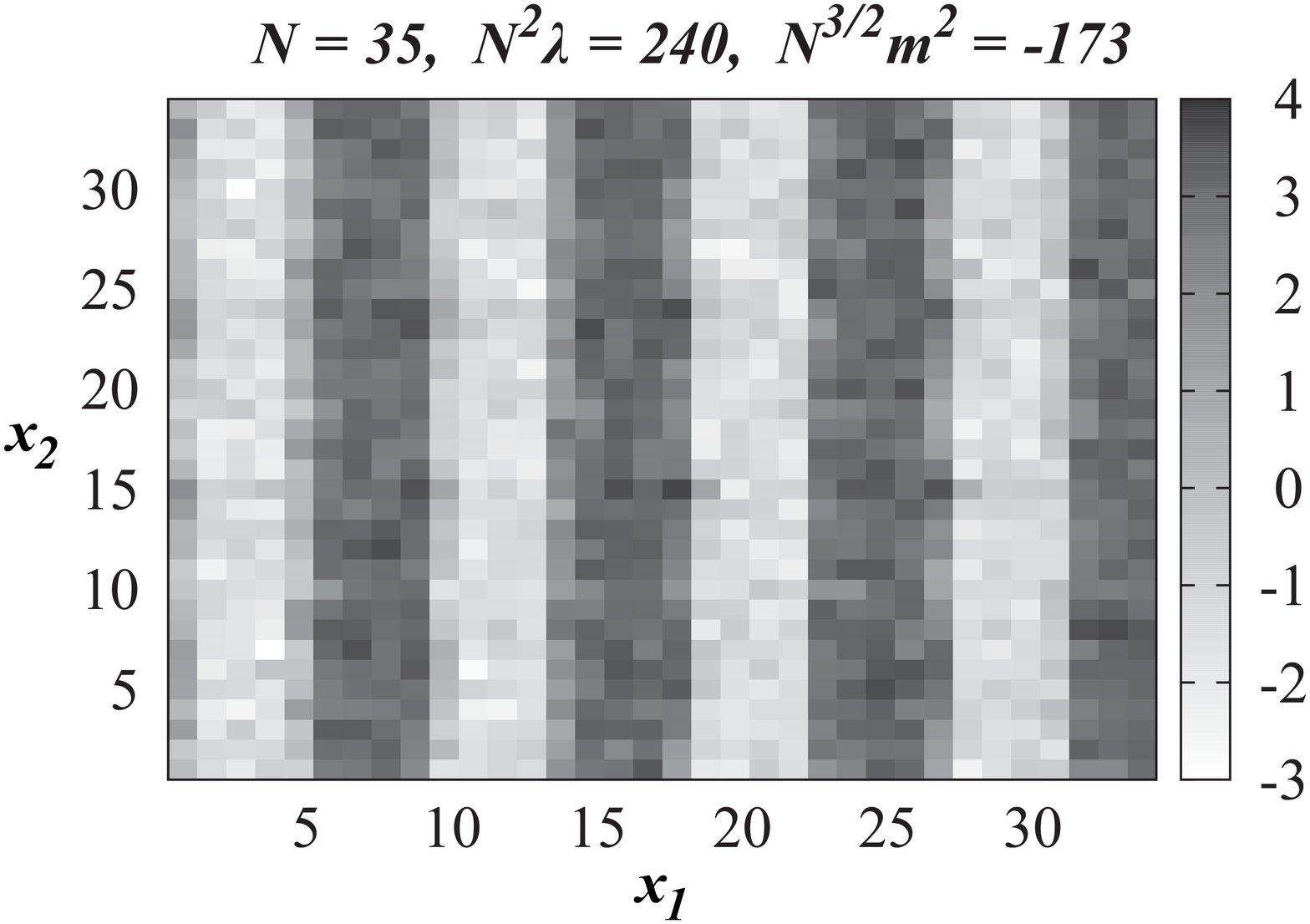}
\caption{An illustration of typical configurations in the four 
sectors of the phase diagram shown in Figure~\ref{phasedia}.
The upper plots are in the disordered phase, next to the uniform
phase (on the left) and to the striped phase (on the right). 
The lower plots show examples of uniform and of striped ordering.}
\label{snap2d}
\end{figure}
The phase diagram in Figure~\ref{phasedia} can be divided into 
four sectors, depending whether $N^{2} \lambda$ and $-N^{3/2}m^{2}$
are small or large. For each of these sectors, 
a typical configuration is shown in Figure~\ref{snap2d}. These
figures are obtained by mapping the $N \times N$ matrices
back to a scalar field configuration, according to the 
prescription in Refs.~\cite{AMNS} (the $\Phi_{x}$ values
correspond to bright or dark spots, according to the colour 
code on the right-hand-side).

\begin{figure}[hbt!]
\center
\includegraphics[angle=0,width=0.5\linewidth]{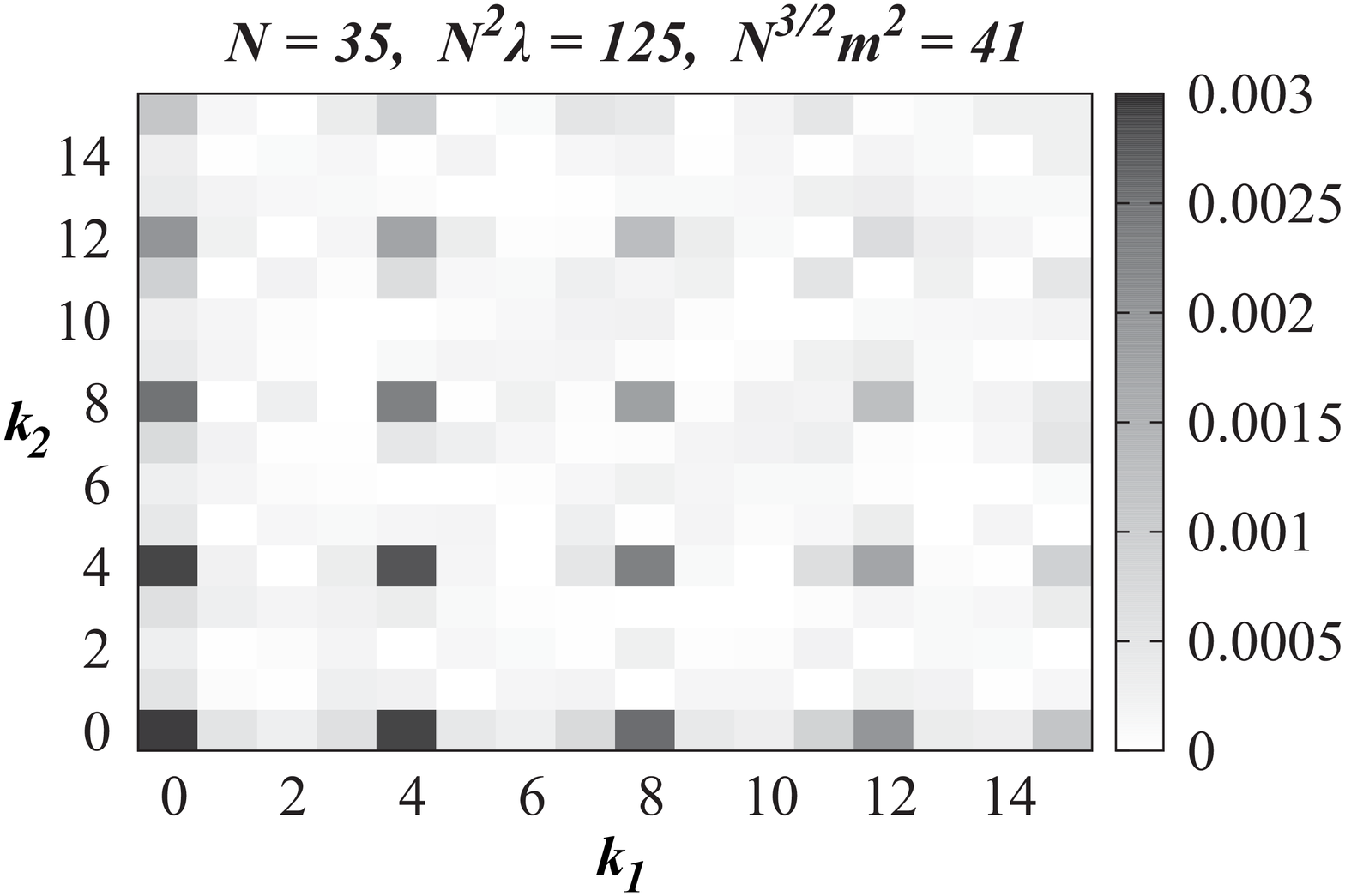}
\hspace{-3mm}
\includegraphics[angle=0,width=0.5\linewidth]{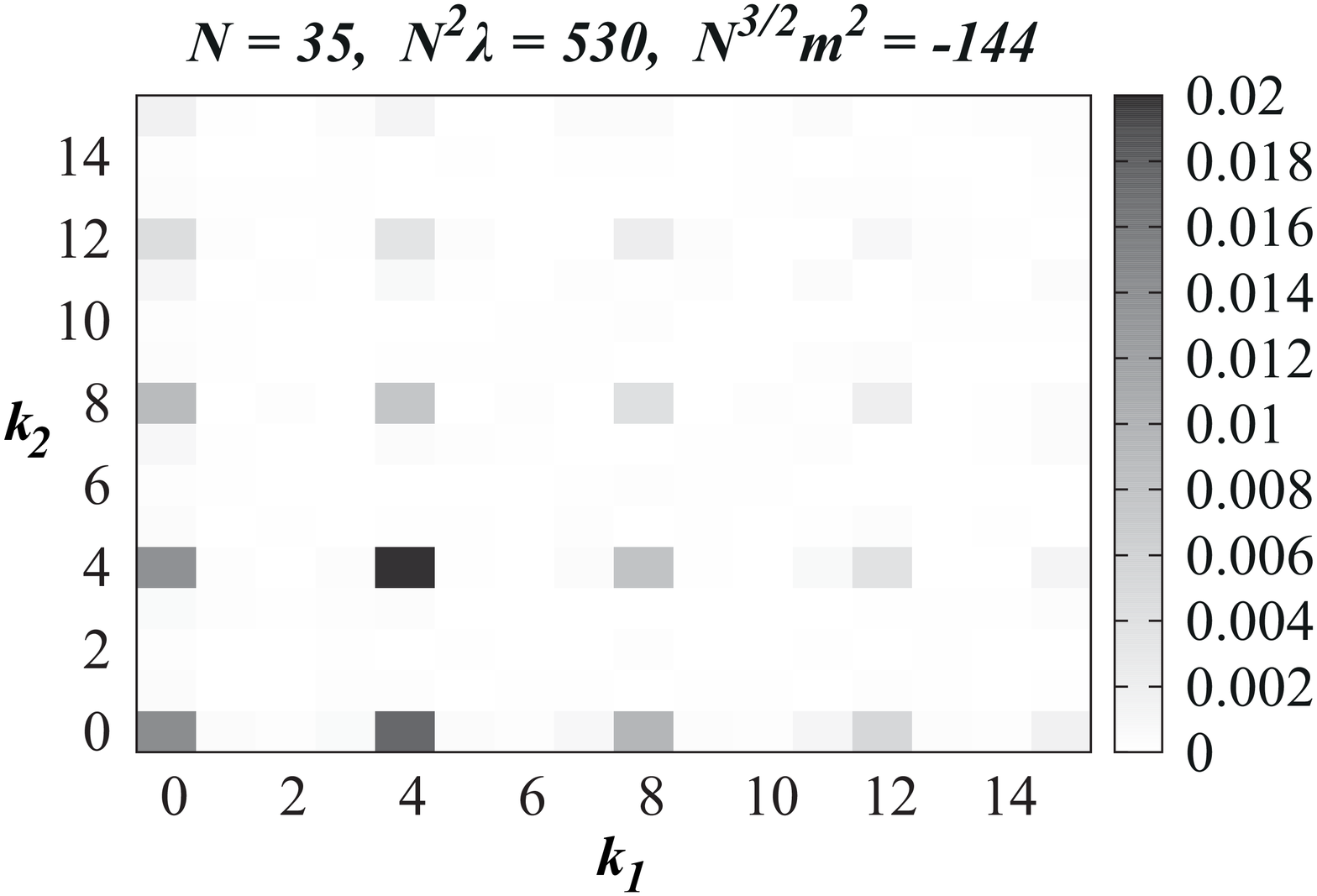}\\
\includegraphics[angle=0,width=0.5\linewidth]{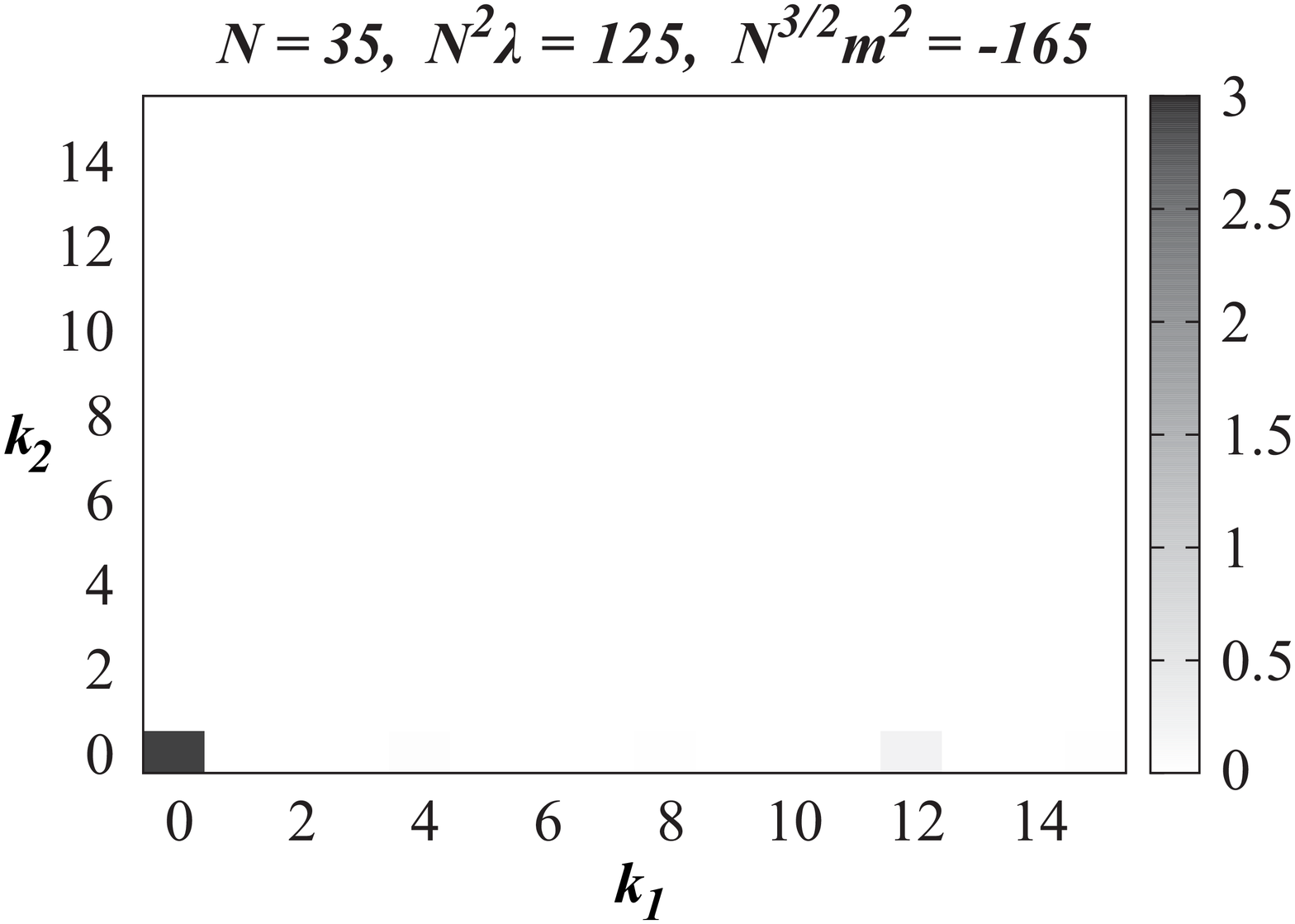}
\hspace{-3mm}
\includegraphics[angle=0,width=0.5\linewidth]{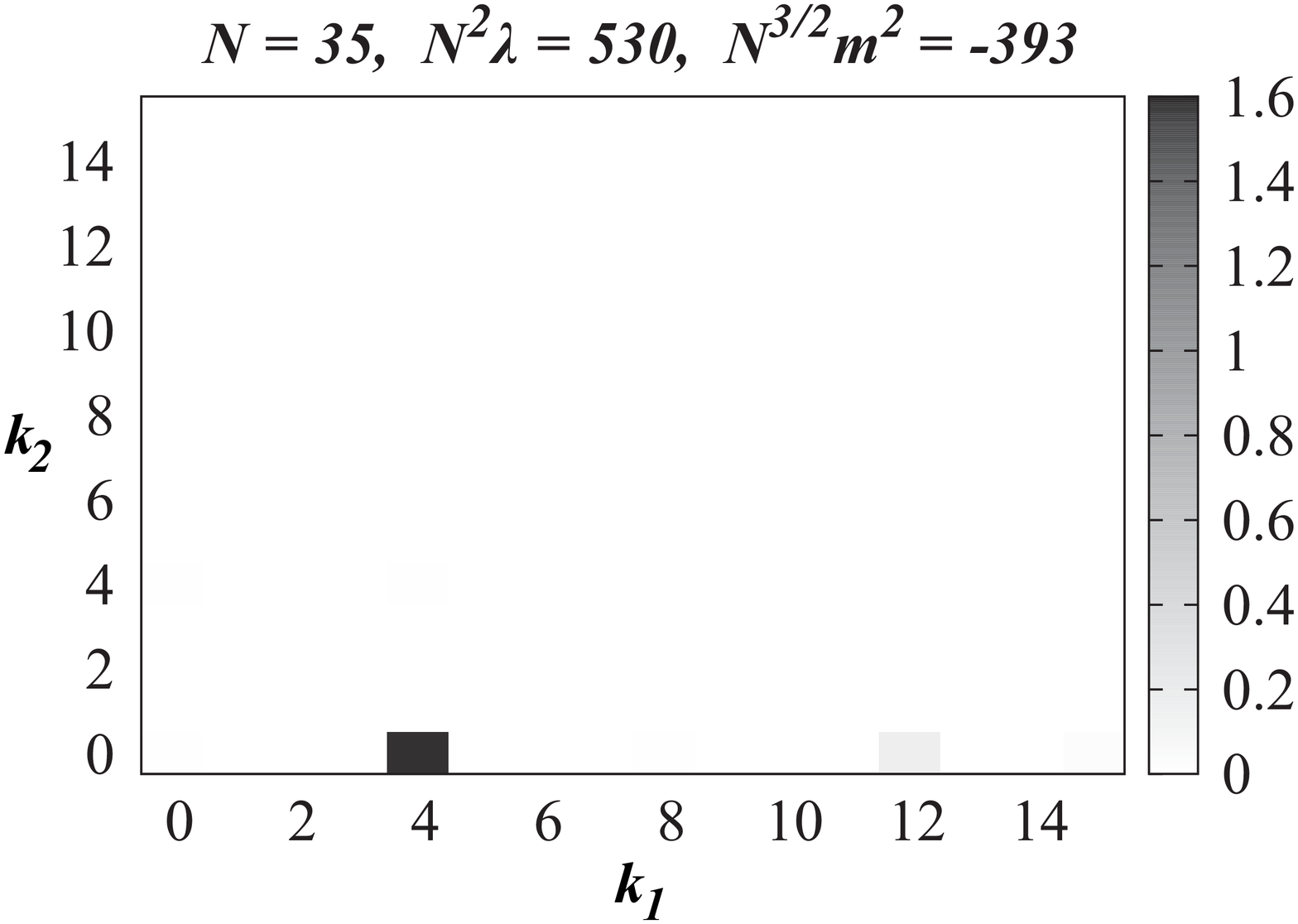}
\caption{Another illustration of typical configurations in the four 
sectors of the phase diagram shown in Figure~\ref{phasedia}.
Here we show typical contributions to the momentum space propagator
$G(\vec k)$ (cf.\ eq.\ (\ref{Gp})) at specific parameters $m^{2}$ and 
$\lambda$. The upper plots are in the disordered phase, next to the 
uniform phase (on the left) and to the striped phase (on the right). 
The lower plots show examples of uniform and of striped ordering.}
\label{propsnap}
\end{figure}

We add an analogous consideration of the propagator in momentum 
space,\footnote{For convenience, the momentum $\vec p$ is re-scaled to
$\vec k = {\textstyle\frac{N}{2\pi}} \vec p$, since the stripe 
patterns correspond to integer components $k_{\mu}$.}
\be  \label{Gp}
G(\vec p ) = \la \tilde \phi (- \vec p ) \, 
\tilde \phi (\vec p ) \ra 
= G \left( {\textstyle\frac{2\pi}{N}} \vec k \right)
\ , \quad \tilde \phi (\vec p) 
= \frac{1}{N^{2}} \sum_{\vec x} \phi_{\vec x} \, e^{-\ri \, \vec p \cdot \vec x} \ .
\ee
The momentum space propagator also played an essential r\^{o}le
in the analysis of Ref.\ \cite{Gubs}. It is worth noting, however, 
that certain properties characterising the propagator in a commutative 
theory do not trivially extend to the NC case. For example, the 
interpretation in terms of a spectral representation is subtle, and the 
meaning of the energy eigenvalues in the case when the non-commutativity
involves the (Euclidean) time coordinate is delicate. However, our
considerations are not affected by these questions.

Figure \ref{propsnap} shows another set of maps in the four
sectors of the phase diagram, where now the brightness/darkness
indicate the value of $G(\vec k )$ in the $(k_{1},k_{2})$ plane.
These maps correspond to the $G(\vec k)$-contributions of typical 
configurations, which fully confirm the picture of Figure \ref{snap2d}.\\

To identify these different phases systematically, we computed the
expectation values of a set of suitable momentum-dependent order 
parameters $M(k)$. For a given absolute value of $k = |\vec k |$, 
and a given field configuration, the value of
$M(k)$ is defined as the maximum of the modulus of the 
Fourier transform of $\Phi_{x}$,
\be \label{ordpar}
M(k) = \frac{1}{N^{2}} \max_{ \frac{N}{2\pi} \left| \vec{p} \right| = k }
\Big\vert \sum_{x}  e^{-i \vec{p} \cdot \vec{x}} \Phi_{x} \Big\vert \ .
\ee
According to this definition, $M(k=0)$ 
reduces to the standard order parameter for uniform magnetisation 
(and the normalisation is such that it equals
the average value of $\Phi_{x}$ on a given configuration). On the
other hand, for non-zero $k$ this observable can detect
a possible dominance of striped patterns (the stripes
are maximally manifest orthogonally to the $\vec k$-direction,
which is summed over in $\la M(k) \ra$). \\

\begin{figure}[hbt!]
\center
\includegraphics[angle=0,width=0.5\linewidth]{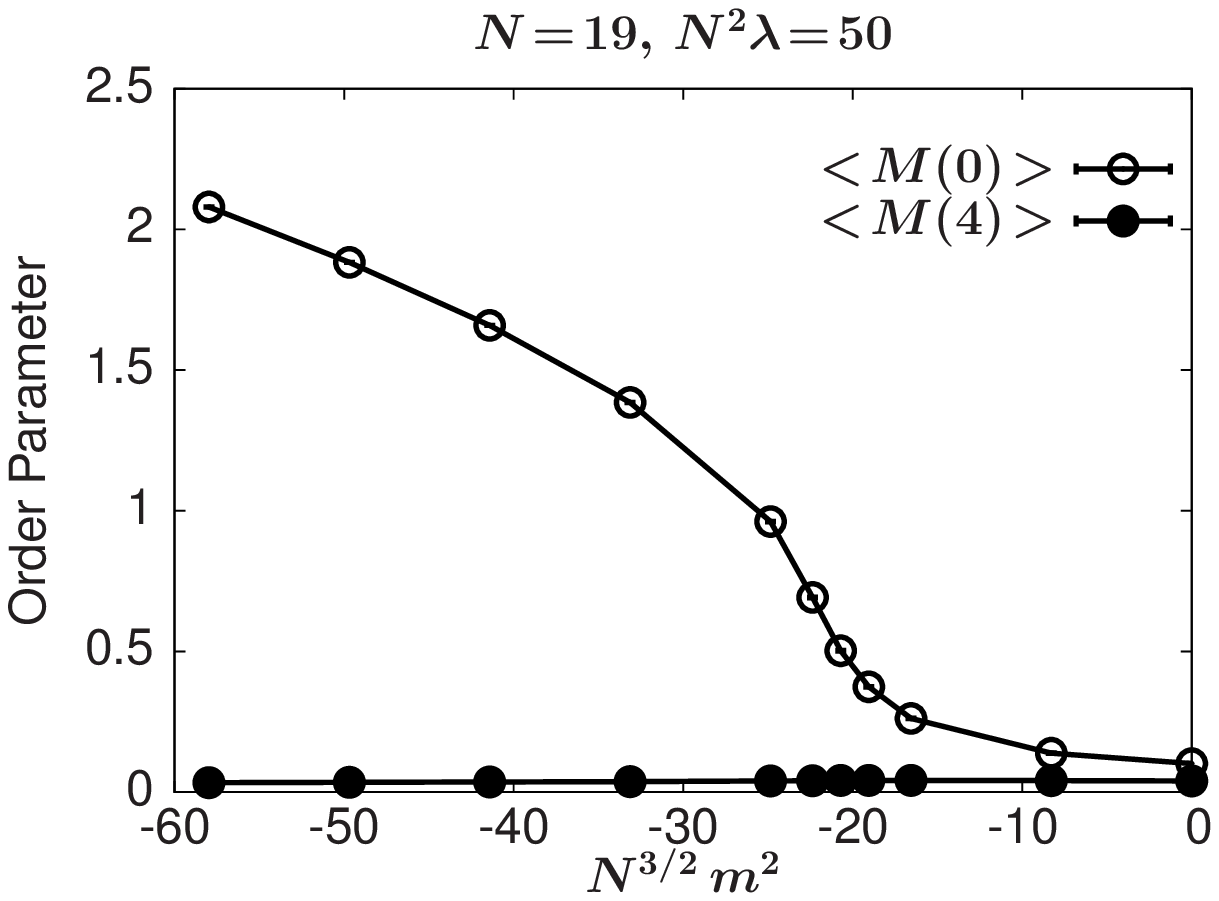}
\hspace{-3mm}
\includegraphics[angle=0,width=0.5\linewidth]{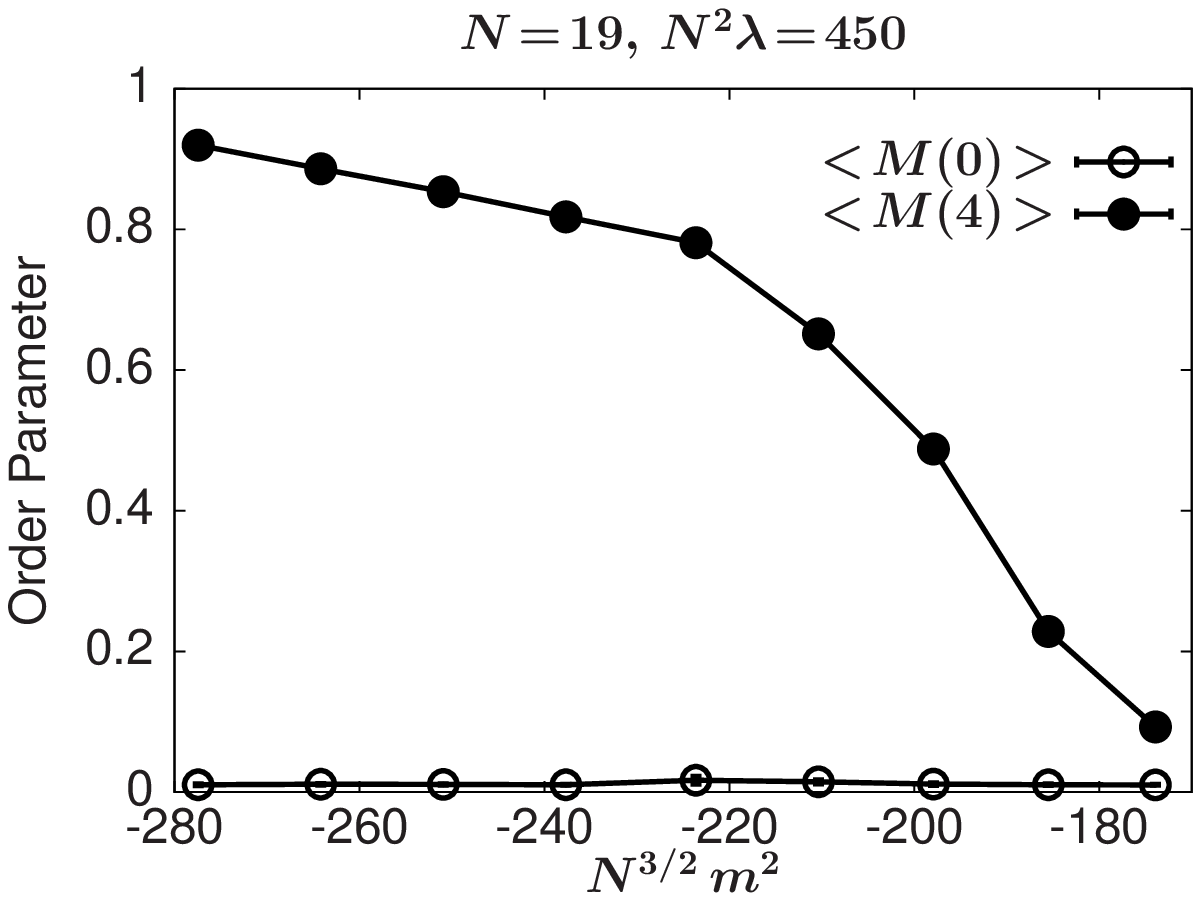}
\caption{The order parameters $\la M(0)\ra$ and $\la M(4) \ra$, which
are defined in eq.\ (\ref{ordpar}). The figures show the dependence on $m^{2}$
at fixed $N$ and $\lambda$. For small (large) $\lambda$ and decreasing 
$m^{2}$, the disorder turns into a uniform (4-stripe) pattern.}
\label{ordparplots}
\end{figure}

Figure \ref{ordparplots} shows examples how the 
order parameters $\la M (0) \ra$ or $\la M(4) \ra$ 
become significant for decreasing $m^{2}$.

\begin{figure}[hbt!]
\center
\includegraphics[angle=0,width=0.51\linewidth]{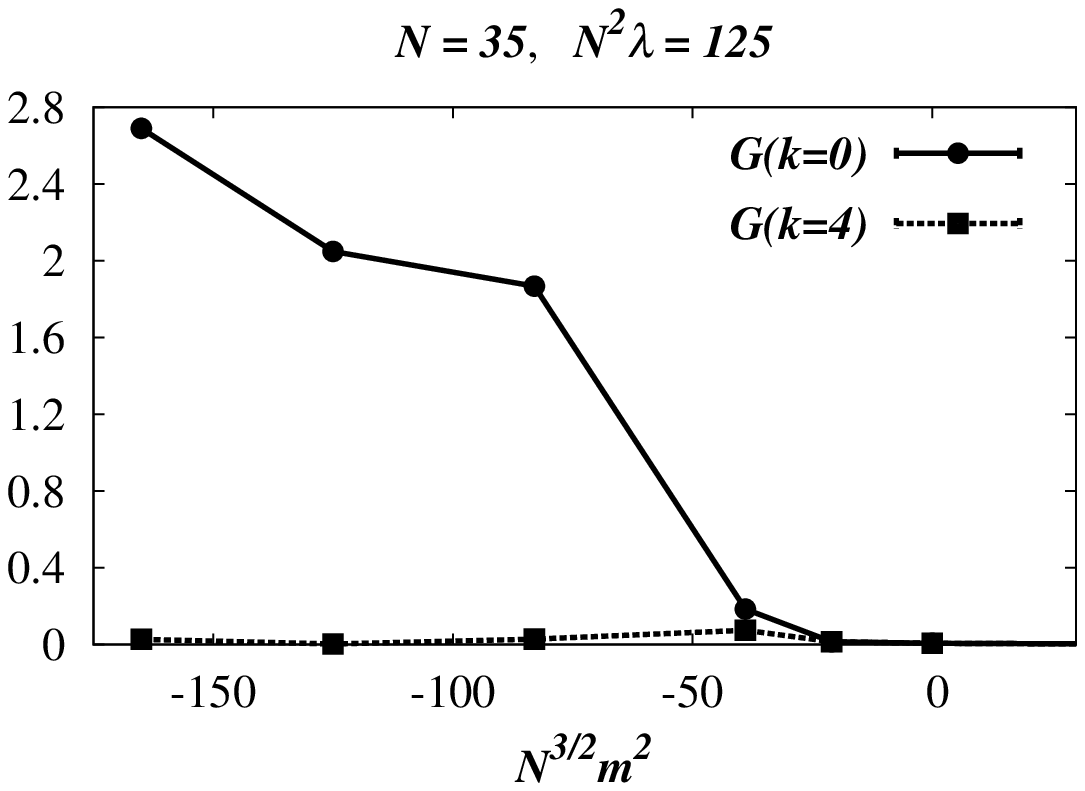}
\hspace{-5.2mm}
\includegraphics[angle=0,width=0.51\linewidth]{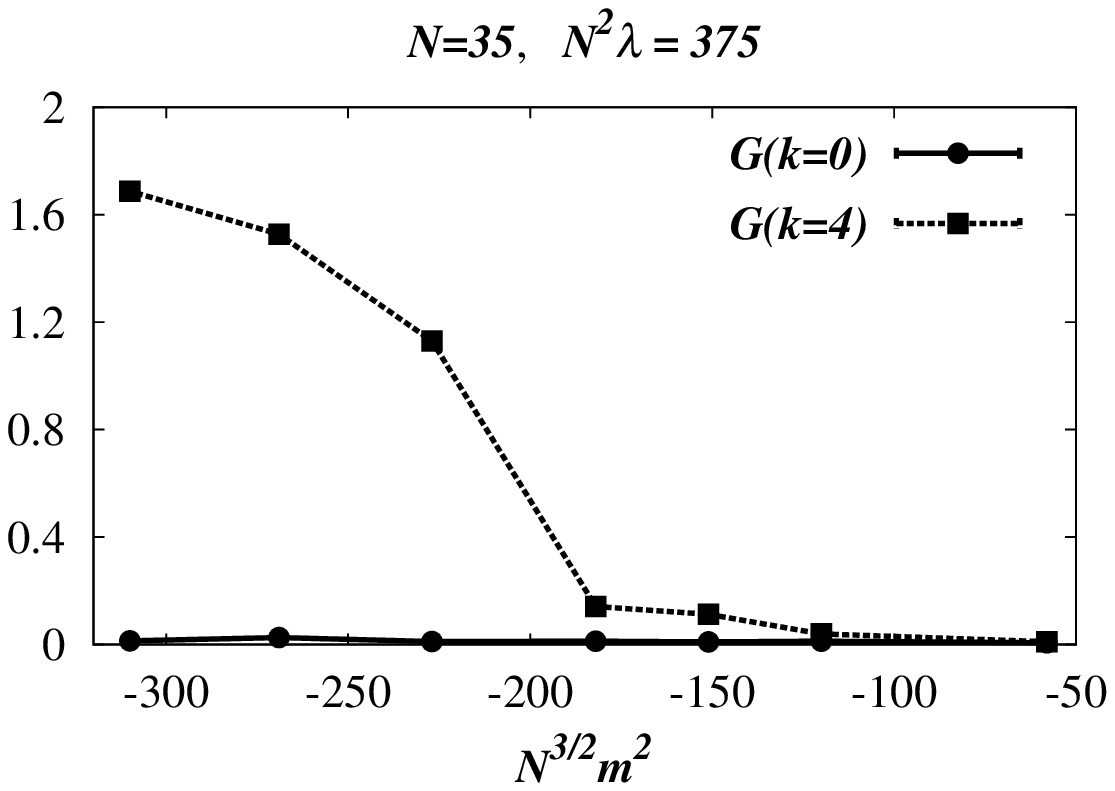}\\
\includegraphics[angle=0,width=0.51\linewidth]{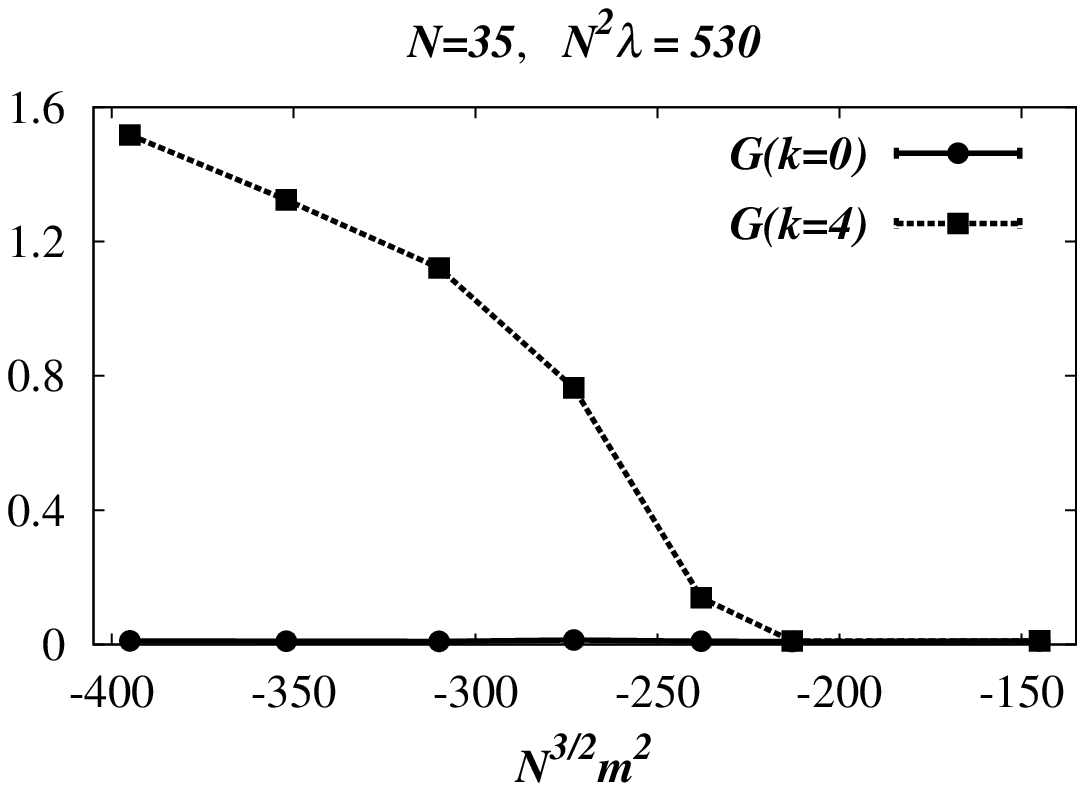}
\hspace{-5.2mm}
\includegraphics[angle=0,width=0.51\linewidth]{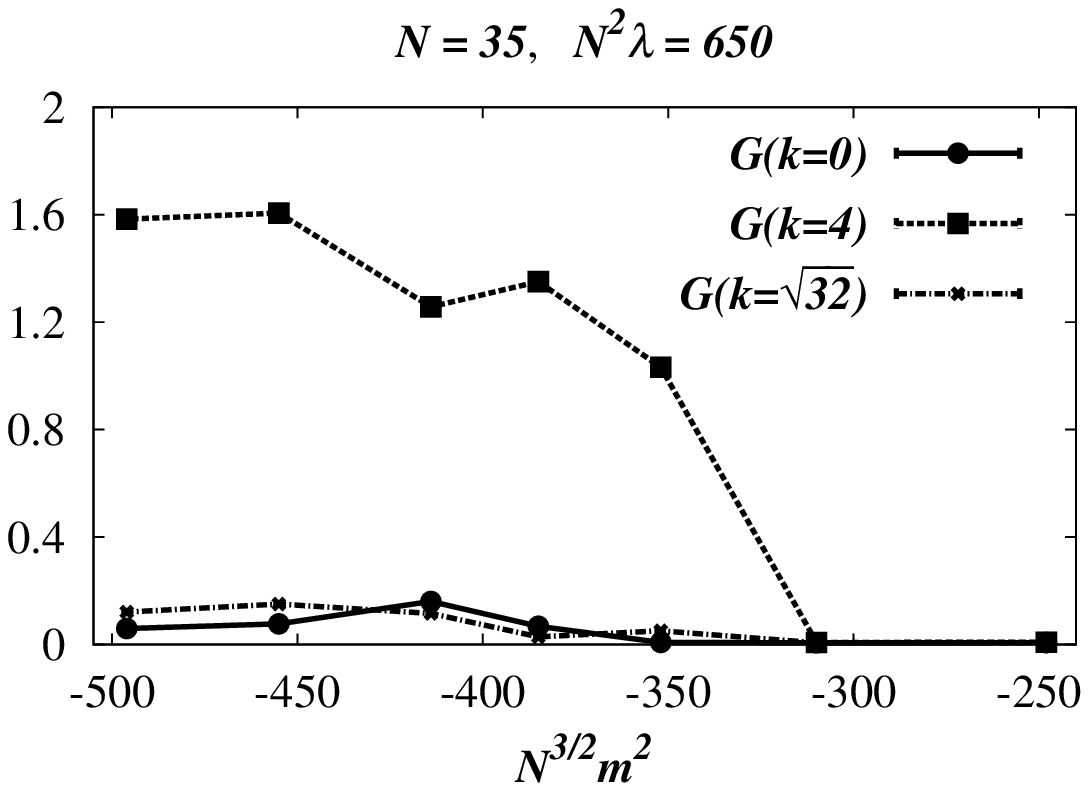}
\caption{The momentum space propagator $G(k)$ (cf.\ eq.\ (\ref{Gp}),
and $k = | \vec k|$) at $N=35$ and fixed $\lambda$, as a function of 
$m^{2}$.
As $m^{2}$ decreases, we confirm the phase transition to an ordered
phase: for small $\lambda$ to the uniform phase (top, left), and
for larger $\lambda$ to the striped phase (remaining three plots).
At large $\lambda$ we also notice a non-negligible contribution
of $G(\vec k = (4,4))$.}
\label{Gpabs}
\end{figure}
Again we add analogous results for the propagator in momentum
space. Figure \ref{Gpabs} shows the expectation values 
$G(k)$ for the relevant values of $k$, as a function of 
$m^{2}$, at fixed $\lambda$. Also here we observe
a behaviour which is entirely consistent with the one of
$\la M(k) \ra$, namely the onset of $G(0,0)$ (of $G(4,0)$
or $G(0,4)$) as $m^{2}$ decreases down to the uniform (striped) 
phase. At the largest $\lambda$-value in this set (plot on the 
bottom, right), we also see some imprint of
$G(4,4)$, {\it i.e.}\ here also the contribution of a 
``checkerboard-type'' pattern is not negligible anymore.\\

\begin{figure}[hbt!]
\center
\includegraphics[angle=0,width=0.47\linewidth]{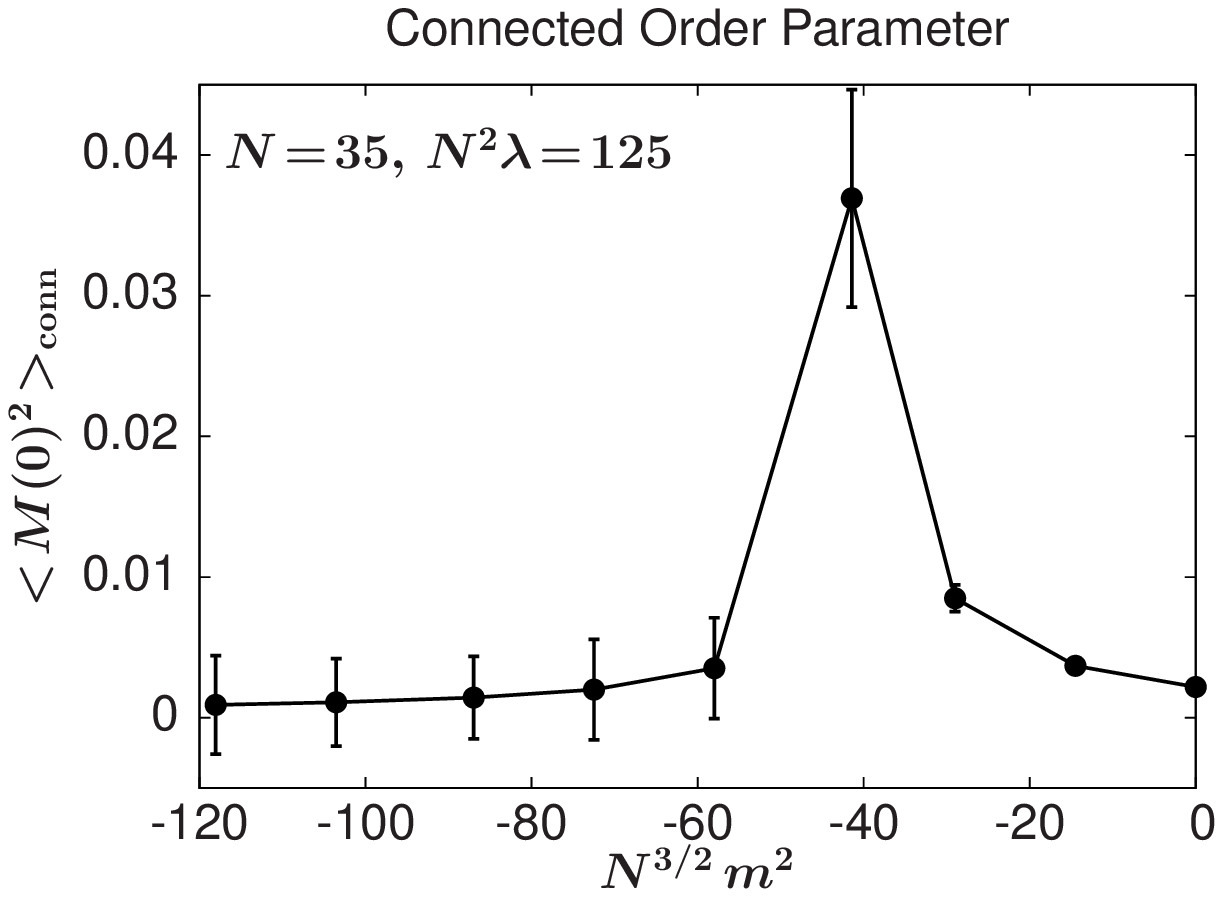}
\hspace{-3mm}
\includegraphics[angle=0,width=0.47\linewidth]{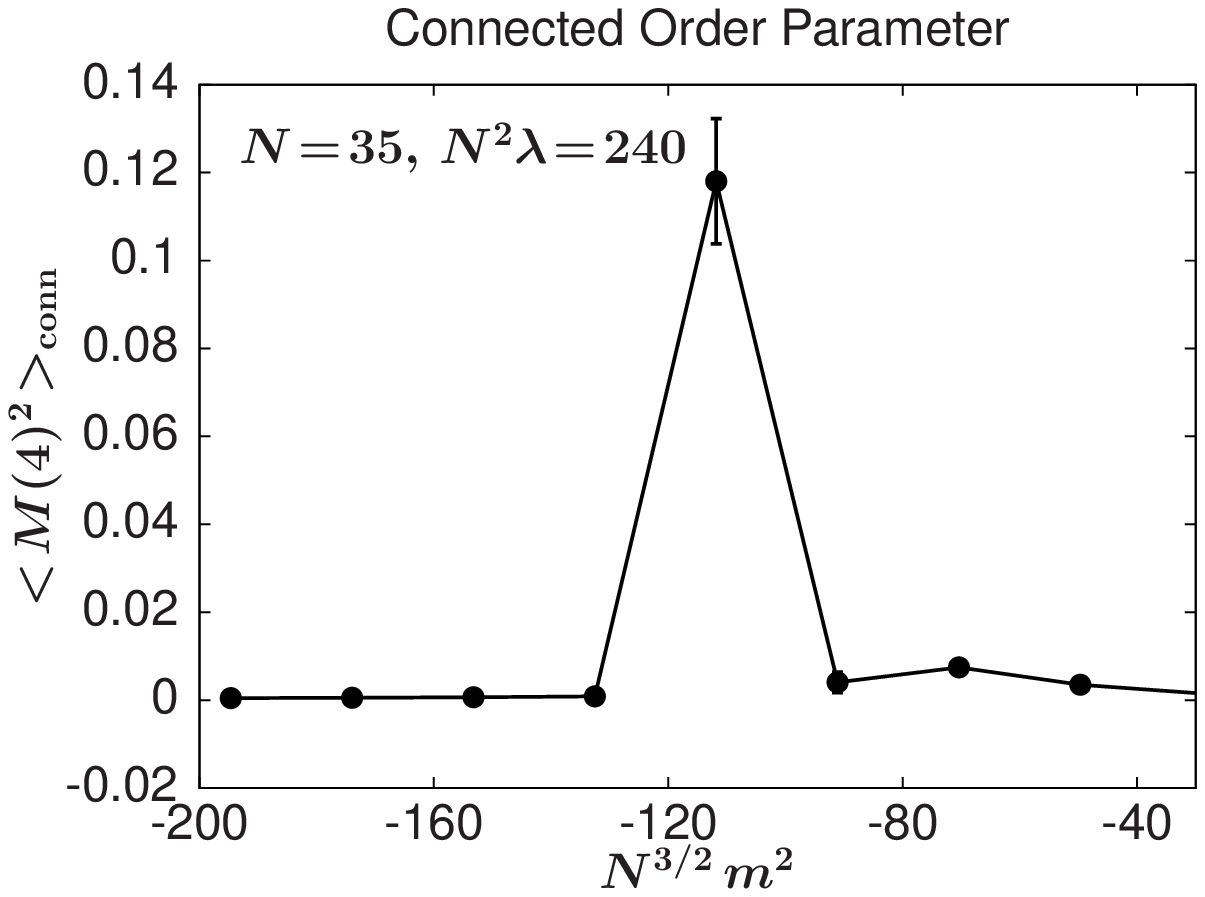}
\caption{The connected correlation function of the order 
parameters $\la M(0)^{2} \ra_{\rm conn}$ and $\la M(4)^{2} \ra_{\rm conn}$. 
In both cases they exhibit a rather sharp peak, which allows us 
to identify the critical value $m^{2}_{\rm c}$.}
\label{ordparacorr}
\end{figure}

Such pictures reveal the type of emerging order unambiguously, 
but for the identification of the critical value $m_{\rm c}^{2}$ it 
is favourable to consider the connected correlation function of 
the relevant order parameter,
\be
\la M (k)^{2} \ra_{\rm conn} = \la M (k)^{2} \ra - \la M (k) \ra^{2} \ .
\ee
Here the transition corresponds to a peak, which can be located 
well, as the examples in Figure \ref{ordparacorr} show. This
property also shows that these order-disorder phase transitions 
are of second order. 

\begin{figure}[hbt!]
\center
\includegraphics[angle=0,width=0.55\linewidth]{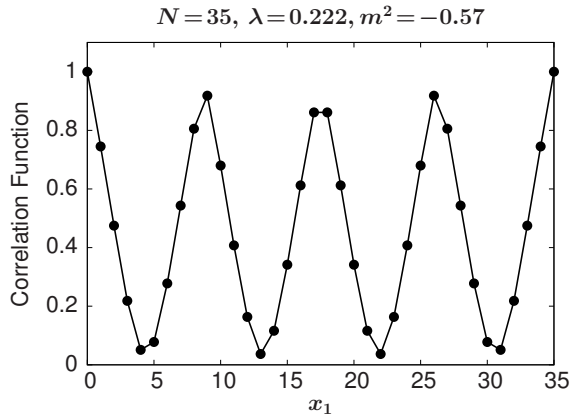}
\caption{The correlation function $\la \phi_{(0,0)} \phi_{(x_{1},0)}\ra$
in the point $(N^{2} \lambda , N^{3/2}m^{2} ) = (272, -118)$, which is
disordered, but close to the transition to the striped phase. 
The figure reveals a trend towards the 4-stripe pattern, which sets
in at somewhat lower $m^{2}$. (We normalised the correlation function
such that $\la \phi_{\vec x} \phi_{\vec x}\ra = 1$.)}
\label{corre}
\end{figure}

At low $\lambda$, this is an Ising-type transition, which
also occurs in the commutative $\lambda \phi^{4}$ model, as argued
in Ref.\ \cite{Gubs}. Regarding the transition to the
striped phase it should be noted, however, that the correlation
function does not decay exponentially, hence it is not possible
to extract a correlation length.
An example for the correlation function in the
disordered phase, but close to striped ordering, is shown
in Figure \ref{corre}. The trend towards a 
4-stripe pattern, which condenses at somewhat lower $m^{2}$, is clearly
visible.

Also in the 3d case non-exponential correlations were observed
within the NC plane \cite{BHN}. However, in that case the decay was 
exponential in the third (commutative) direction, which
did provide a correlation length, along with the aforementioned
dispersion relation $E (\vec p)$.

%% file: DSL2.tex
So far, the discussion of our results has been in terms of 
quantities expressed in lattice units.
In order to discuss the DSL, however, it is necessary to introduce a
dimensional quantity to ``set the scale'' and to be able to
take the continuum limit. In particular, a question of central relevance 
in our analysis is whether the DSL can be taken while remaining in the 
proximity of the striped phase: if this is the case, it provides a piece 
of evidence for the persistence of that phase in the continuum theory
at fixed non-commutativity parameter.

Since the correlation function does not decay exponentially, 
it is not possible to resort to the standard procedure, which refers 
to the correlation length as the natural ``physical'' scale.
Nevertheless, an appropriate physical scale can be extracted 
from the decay of the correlation function, which becomes a slowly 
varying function of the distance when $m^{2}$ approaches $m^{2}_{c}$.
To suppress finite size effects, we consider the limit
\be
\Delta m^{2} := m^{2} - m^{2}_{\rm c} \, \to \, 0
\ee
from above, {\it i.e.}\ within the disordered phase. The goal 
is to increase $N$ at the same time, in such a way that the 
correlation function remains stable down to the first dip.
Thus $\Delta m^{2}$ will be converted into the factor needed
for the DSL, with a critical exponent that we denote as $\sigma$,
\be \label{am2}
a^{2} \propto (\Delta m^{2})^{\sigma} \ ,
\ee
and which remains to be determined.

As a normalisation, we choose the units such that $a=1$
at $N=35$, which means that distances take the form
\be  \label{ax}
a \vec x = \sqrt{\frac{35}{N}} \vec x \ ,
\ee
if we keep $N a^{2}$ fixed, as required for the DSL.
We relate this term to $\Delta m^{2}$ as
\be \label{ansatz}
N a^{2} = N \frac{(\Delta m^{2})^{\sigma}}{(m_{\rm c}^{2})^{1+ \sigma}} \ .
\ee
This is an ansatz, which is consistent with the proportionality
relation (\ref{am2}), and which expresses the distance from the continuum
limit in terms of a dimensionless ratio (much in the
spirit of the dimensionless temperature $\tau := (T - T_{\rm c})/T_{\rm c}$, 
or the dimensionless pressure, often used to parameterise the 
vicinity of a phase transition in thermal statistical mechanics). 
More explicitly, the right-hand side of eq.~(\ref{ansatz})
accounts for the fact that $N a^{2}$ has energy dimension $-2$, 
so the suppression factor to the trivial 
proportionality relation $N a^{2} \propto N/m_{\rm c}^{2}$ depends
on a dimensionless ratio, like $\Delta m^{2}/m_{\rm c}^{2}$.

\begin{figure}[hbt!]
\center
\includegraphics[angle=0,width=0.48\linewidth]{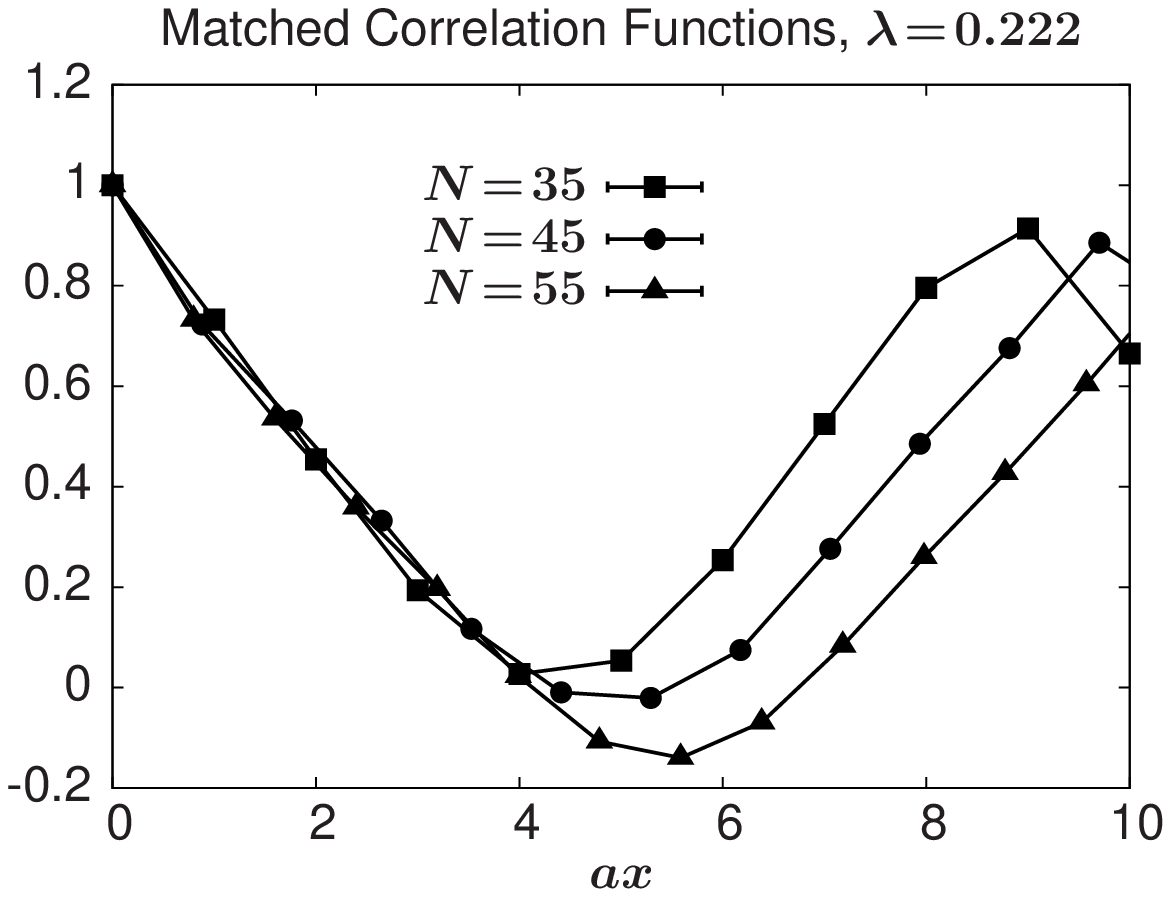} \vspace*{3mm} \\
\includegraphics[angle=0,width=0.48\linewidth]{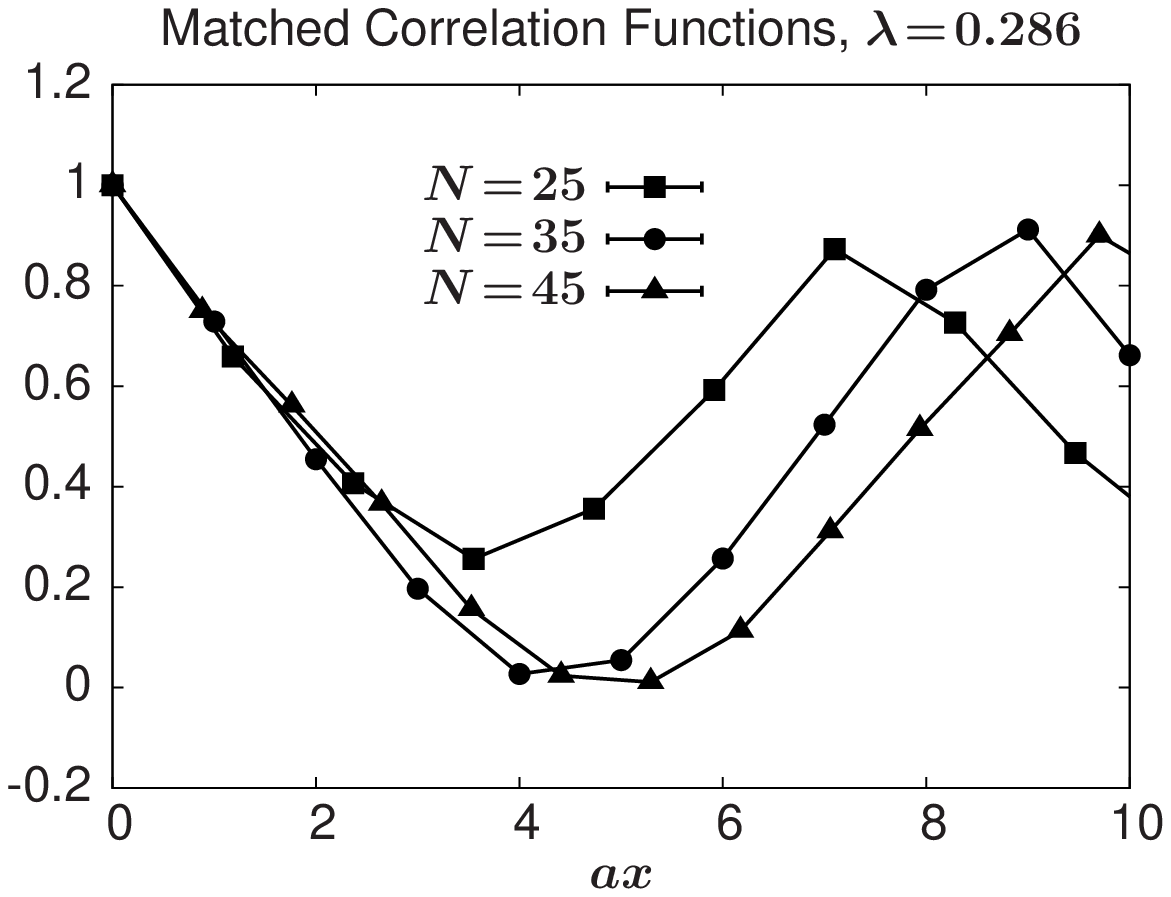} \vspace*{3mm} \\
\includegraphics[angle=0,width=0.48\linewidth]{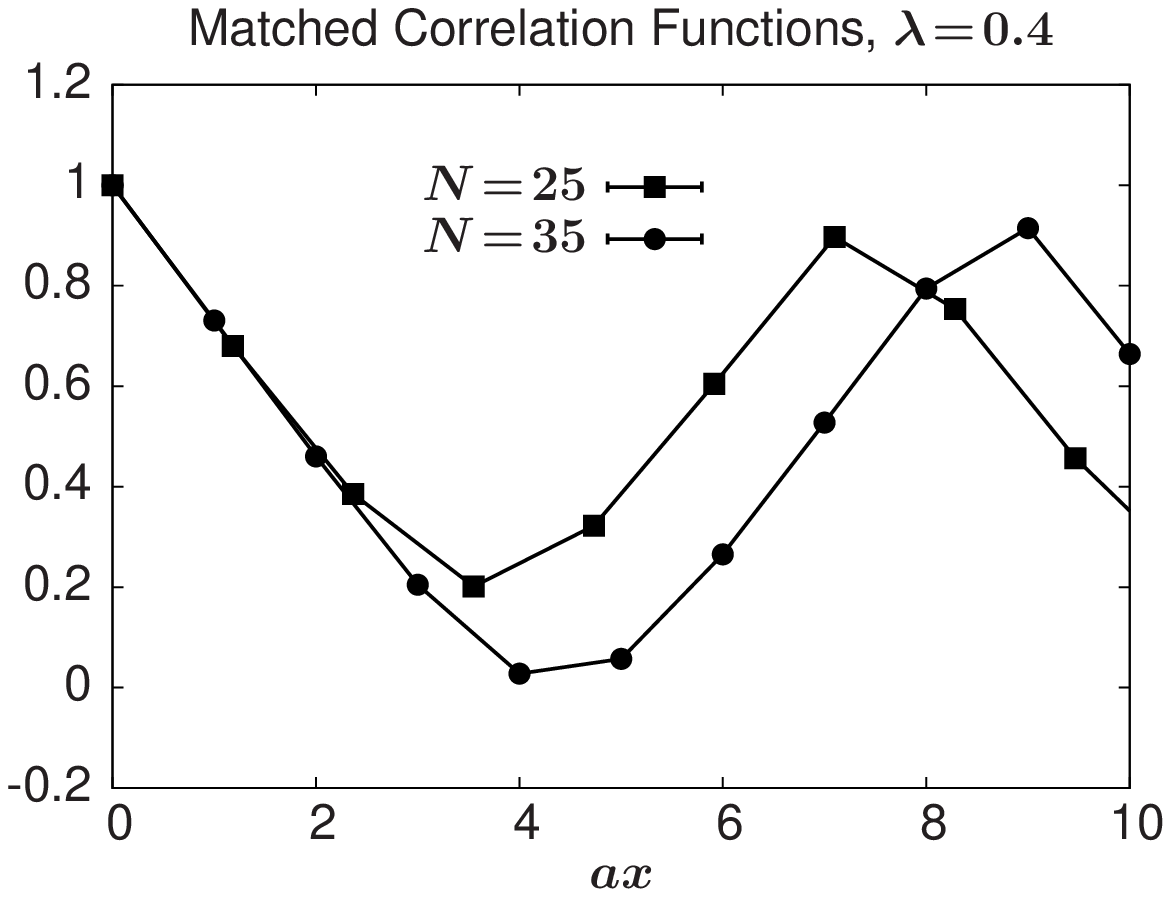}
\caption{Examples of ``matched correlation functions'': 
at different system sizes $N$ and at fixed coupling $\lambda$, 
$\Delta m^{2} = m^{2} - m^{2}_{\rm c}$ is tuned such that
the short-distance correlations agree. Then the distance in 
physical units --- as given in eq.\ (\ref{ax}) --- agrees as well. Thus 
we identify the $\Delta m^{2}$ values to be inserted in eq.\ (\ref{sig}),
which fixes the critical exponent $\sigma$. (Again the correlation
functions are normalised to $1$ at $x =| \vec x | = 0$.)}
\label{corrematch}
\end{figure}

The practical method to determine $\sigma$ proceeds as follows:
we consider two sizes $N_{1}$ and $N_{2}$, and we search for the 
parameters which correspond to the same trajectory towards the 
DSL. Hence we fix the same coupling $\lambda$, so that the 
dimensionless product $\lambda \theta$ is kept constant.
Then we adjust values of $\Delta m_{1}^{2}$ and $\Delta m_{2}^{2}$,
in such a manner that the correlation decays coincide 
(as well as possible) until the first dip is reached.
Figure \ref{corrematch} shows examples for such ``matched
correlations functions'' at three values of $\lambda$, and
various sizes in each case.

Having identified such pairs $\Delta m_{1}^{2}$, $\Delta m_{2}^{2}$
--- and the corresponding critical values $m_{1,{\rm c}}^{2}$, 
$m_{2,{\rm c}}^{2}$ --- we extract the critical exponent
\be  \label{sig}
\sigma = \frac{\ln (m_{1,{\rm c}}^{2} / m_{2, {\rm c}}^{2})}
{\ln (\Delta m_{1}^{2} / \Delta m_{2}^{2})
+ \ln (m_{1,{\rm c}}^{2} / m_{2,{\rm c}}^{2})} \ .
\ee
This can be done for various pairs $N_{1}$, $N_{2}$, at fixed
$\lambda$, always in the vicinity of the striped phase.
In practice, the accessible values of $N^{2} \lambda$
are restricted by the feasibility of the simulation:
for $N_{1} < N_{2}$ the value of $\lambda$ has to be large 
enough for $N_{1}^{2} \lambda$ to be close to the striped phase,
but $N_{2}^{2} \lambda$ should not become too large, to avoid
a situation with a multitude of deep meta-stable minima,
where the Monte Carlo history could get stuck. On the other
hand, $N_{1}$ and $N_{2}$ should differ significantly ---
and both should be large enough to attain the asymptotic 
large-volume regime ---
for the results for the $\sigma$ exponents to be sensible.

\begin{figure}[hbt!]
\centering
\includegraphics[angle=0,width=0.65\linewidth]{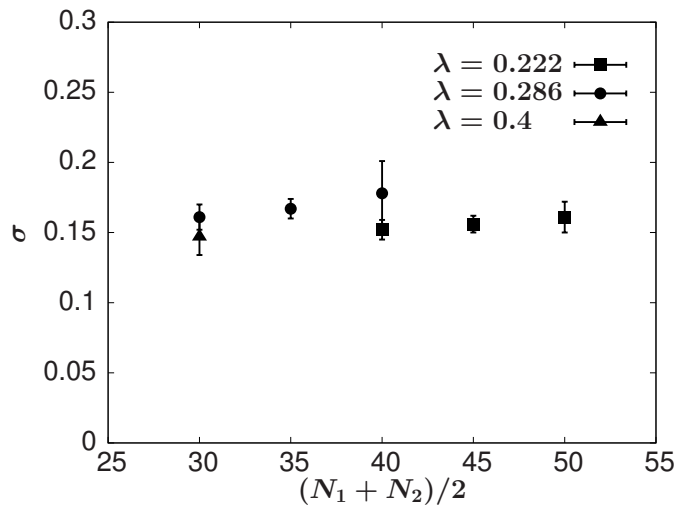}
\caption{An illustration of the values given in Table \ref{tabsigma}.
There is a clear trend to a plateau value of $\sigma = 0.16(1)$.}
\label{sigmaplot}
\vspace*{-2mm}
\end{figure}

If these results stabilise for increasing $N_{1}$, $N_{2}$
and decreasing $\Delta m_{1}^{2}$, $\Delta m_{2}^{2}$, 
we can conclude that it is indeed possible to take a DSL
next to the striped phase, so that the latter persists ---
otherwise it would likely be removed in the DSL.
Table \ref{tabsigma} and Figure \ref{sigmaplot} show our results, 
which explore the window of sizes and couplings, which are 
numerically well tractable. 
\begin{table}[h!]
\centering
\begin{tabular}{|c||c|c|c||c|c|c||c|}
\hline
$\lambda$ & $N_{1}$ & $m^{2}$ & $m_{\rm c}^{2}$ & $N_{2}$ & 
$m^{2}$ & $m_{\rm c}^{2}$ & $\sigma$ \\ 
\hline
\hline
\multirow{3}{*}{0.222} & 35 & \multirow{2}{*}{$-0.59$} & 
\multirow{2}{*}{$-0.60(3)$} & 45 & $-0.77$ & $-0.83(4)$ & 0.152(7) \\
 & 35 & & & 55 & \multirow{2}{*}{$-0.84$} & \multirow{2}{*}{$-0.99(4)$} 
 & 0.156(6) \\
 & 45 & $-0.77$ & $-0.83(4)$ & 55 & & & ~ 0.161(11) \\
\hline
\multirow{3}{*}{0.286} & 25 & \multirow{2}{*}{$-0.59$} & 
\multirow{2}{*}{$-0.63(2)$} & 35 & $-0.66$ & $-0.85(3)$ & 0.161(9) \\
 & 25 & & & 45 & \multirow{2}{*}{$-0.57$} & \multirow{2}{*}{$-1.03(5)$} 
 & 0.167(7) \\
 & 35 & $-0.66$ & $-0.85(3)$ & 45 & & & ~ 0.178(23) \\
\hline
 0.4 & 25 & $-0.88$ & $-0.95(5)$ & 35 & $-0.91$ & $-1.25(7)$ 
& ~ 0.147(13) \\
\hline
\end{tabular}
\caption{The $\sigma$-values obtained for various pairs of
sizes $N_{1}$, $N_{2}$ after tuning $\Delta m^{2}$ such that the
short-distance decay of the correlation functions coincide.}
\label{tabsigma}
\end{table}

We see a convincing trend towards a stable critical exponent 
\be
\sigma = 0.16(1) \ .
\ee
Therefore our results give evidence for the 
persistence of the striped phase in the DSL.

%% file: conclu2.tex
We have presented a non-perturbative, numerical
study of the $\lambda \phi^{4}$ model on a NC plane.
Regularising the model as a lattice theory, and mapping
the latter to a matrix formulation, enabled us to determine
the phase structure of this model by means of Monte Carlo simulations, 
without relying on weak- or strong-coupling assumptions. 

Our study of the phase diagram of the lattice model at
finite lattice spacing reveals the existence of phases
corresponding to disorder, uniform order and non-uniform
order, in agreement with expectations based on analytical
studies, and with previous numerical work on this and in
related models. In particular, the typical configurations
in the non-uniform order phase are characterised by spontaneous
breaking of rotational and translational symmetries, with
patterns of separated regions, in which the field takes different
values (``stripes''). This phase is absent in the commutative case.

The three different phases are separated by transition lines, which
stabilise rapidly when the system size $N$ is increased, 
if the axes are scaled suitably --- that prescription differs
from the 3d case.

In order to define a ``physical'' value for the lattice
spacing $a$ and take the continuum limit (at fixed non-commutativity
parameter), we have introduced a dimensional scale
based on the decay of the correlation function
(although this decay is not exponential). This allows us
to identify trajectories in the parameter space, which
can be extrapolated to the DSL, {\it i.e.}\ to the continuum 
and to infinite volume, at a constant NC parameter $\theta$.

We have provided evidence that a DSL can be taken in the
vicinity of the striped phase, which means that this
exotic phase does persist in the continuum and in infinite
volume. This implies in particular that the spontaneous
breaking of translation and rotation symmetry occurs.
The apparent contradiction with the Mermin-Wagner theorem
is avoided by the fact that this model is non-local, and
that its IR behaviour is not smooth. \\

%% file: ackno.tex
\noindent
{\small
{\bf Acknowledgements} \ We are indebted to Frank Hofheinz for
his help with the design of the figures. We also thank him, as well
as Jan Ambj\o rn, Antonio Bigarini, Steven Gubser, Jun Nishimura, 
Denjoe O'Connor and Jan Volkholz for useful communications.
This work was supported by the Mexican {\it Consejo Nacional de Ciencia 
y Tecnolog\'{\i}a} (CONACyT) through project 155905/10 ``F\'{\i}sica 
de Part\'{\i}culas por medio de Simulaciones Num\'{e}ricas'',
and by the Spanish {\it MINECO} (grant FPA2012-31686 and 
``Centro de Excelencia Severo Ochoa'' programme grant 
SEV-2012-0249). The simulations were performed on the cluster 
of the Instituto de Ciencias Nucleares, UNAM.
}

%% file: numtech.tex
In the formulation that we simulated, a configuration is given 
by a Hermitian $N \times N$ matrix $\Phi$, where $N$ is odd,
cf.\ Section 2. We applied the standard Metropolis algorithm, and 
proceeded with local, Hermiticity-preserving, updates of the matrix elements,
$\Phi_{ij} \to \Phi_{ij}'$, $\Phi_{ji} \to \Phi_{ij}^{' \ *}$.

In this appendix we comment on a numerically efficient
treatment of the action (\ref{matmod}),
\bea
S[\Phi ] &=& N \ {\rm Tr}\ \left[ \frac{1}{2} \sum_{\mu =1}^{2}
( \Gamma_{\mu} \Phi \Gamma_{\mu}^{\dagger} - \Phi )^{2}
+ \frac{\bar m^{2}}{2} \Phi^{2} + \frac{\bar \lambda}{4} \Phi^{4} 
\right] \nn \\
&:=& N \left[ \frac{1}{2} (s_{\rm kin,1} + s_{\rm kin,2}) + 
\frac{\bar m^{2}}{2} s_{m} + \frac{\bar \lambda}{4} s_{\lambda} \right] \ .
\eea
The twist eaters $\Gamma_{\mu}$ are given in Section 2; note that
the diagonal elements of $\Gamma_{2}$ are powers of
$z = - e^{- \, \ri \, \pi /N}$.

To discuss the evaluation of these terms, we mention that
any Hermitian matrix $H$ fulfils
\be  \label{Htrace}
{\rm Tr} \, H^{2} = \sum_{ij} |H_{ij}|^{2} =
\sum_{i} H_{ii}^{2} + 2 \sum_{i > j} |H_{ij}|^{2} \ .
\ee

$\bullet$ We first address $s_{\rm kin,1}$, where it is useful
to introduce the notation
\be
\bar i := i \ {\rm mod} \ N \ 
\ee
The matrix $\Gamma_{1} \Phi \Gamma_{1}^{\dagger} - \Phi$
is also Hermitian, hence we apply the identity (\ref{Htrace}) 
to arrive at the computationally economic form
\be
s_{\rm kin,1} = \sum_{i} ( 
\Phi_{\overline{i+1},\ \overline{i+1}} - \Phi_{ii})^{2} 
+ 2 \sum_{i>j} | \Phi_{\overline{i+1},\ \overline{j+1}}
- \Phi_{ij}|^{2} \ .
\label{kin1}
\ee

$\bullet$ For the $s_{\rm kin,2}\,$ term, note that
\be
(\Gamma_{2} \Phi \Gamma_{2}^{\dagger})_{ij} =  \Phi_{ij} z^{i-j}
\ee
is still Hermitian, and 
$\Gamma_{2} \Phi \Gamma_{2}^{\dagger} - \Phi$ as well. Inserting
$z$ yields
\bea
s_{\rm kin,2} &=& \sum_{ij} \left| \Phi_{ij} \, ( z^{i-j} -1)\right|^{2} \nn \\
&=& 4 \sum_{i>j} | \Phi_{ij} |^{2} \ 
\left[ 1 - (-1)^{i-j} \cos \frac{\pi (i-j)}{N} \right] \ . \qquad
\label{skin2}
\eea

$\bullet$ There are also simplifications of the action difference, 
which is needed in the Metropolis accept/reject step. For 
\be
\Delta s_{\rm kin} := s_{\rm kin,1}[\Phi '] + s_{\rm kin,2}[\Phi '] -
s_{\rm kin,1}[\Phi ] - s_{\rm kin,2}[\Phi ] \ .
\ee
we obtain
\bea
\Delta s_{\rm kin}(i=j) 
&=& 2 \Big[ \Phi_{ii}^{' ~ 2} - \Phi_{ii}^{2} + (\Phi_{ii} - \Phi'_{ii}) 
\, (\Phi_{\overline{i+1},\ \overline{i+1}} +
\Phi_{\overline{i-1},\ \overline{i-1}} ) \Big] \nn \\
\Delta s_{\rm kin}(i \neq j) 
 &=& 4 \Big[ ( | \Phi_{ij}'|^{2} - | \Phi_{ij}|^{2} ) \,
\Big( 2 - \cos \frac{(N+1) \pi (i-j)}{N} \Big) \nn \\
&& + \ {\rm Re} \ \Big( (\Phi_{ij} - \Phi'_{ij}) \, 
( \Phi_{\overline{i+1},\ \overline{j+1}} +
\Phi_{\overline{i-1},\ \overline{j-1}} ) \Big) \Big] \ . 
\label{Dskin}
\eea
The cosine function in this formula, and in eq.\ (\ref{skin2}),
is only required for $(N-1)$ different arguments, which should be
stored in a look-up array.\\

$\bullet$ The {\em mass term} $s_{m} = {\rm Tr} \, \Phi^{2}$ is 
quick to  evaluate, thanks to identity (\ref{Htrace}).
The main computational challenge is the {\em quartic term}
\be  \label{diffsl}
s_{\lambda} = {\rm Tr} \, \Phi^{4} = \sum_{r} (\Phi^{2})_{rr}^{2}
+ 2 \, \sum_{r>s} | (\Phi^{2})_{rs}|^{2} \ .  
\ee

When updating a generic matrix element $\Phi_{ij}$, $i \geq j$, 
the difference $\Delta s_{\lambda}$ is affected by the 
following elements of $\Phi^{2}$,
\be  \label{phi2rs}
(\Phi^{2})_{is} \ \ {\rm for} \ \ s = i \dots N \ , \quad
(\Phi^{2})_{rj} \ \ {\rm for} \ \ r = 1 \dots j \ .
\ee
These elements have to be computed explicitly for $\Phi^{2}$ 
and for $\Phi^{' \, 2}$.
A final detail is that the elements for the same indices
$r,s$ --- belonging to the set specified in (\ref{phi2rs}) ---
can be obtained without doing the full summation twice,
since most contributions are identical,
\bea
(\Phi^{' \, 2})_{rs} &=& (\Phi^{2})_{rs} + 
\left\{ \begin{array}{cccc}
|\Phi_{ij}'|^{2}- |\Phi_{ij}|^{2} && {\rm if} & r=s \\
(\Phi_{ij}'- \Phi_{ij}) \Phi_{js} && {\rm if} & i=r \neq s \\
\Phi_{ri} (\Phi_{ij}'- \Phi_{ij}) && {\rm if} & j = s \neq r \\
\end{array} \right. \ \ .
\eea